\documentclass[journal=jacsat,manuscript=article]{achemso}
\usepackage[breaklinks=true,colorlinks,citecolor=blue,linkcolor=blue,urlcolor=blue]{hyperref}
\usepackage[table,xcdraw]{xcolor}
\usepackage{epsfig, mathrsfs, color, latexsym, subfigure, marginnote, gensymb, amsmath}
\usepackage{graphicx}
\usepackage{xcolor}
\usepackage{booktabs}
\usepackage{adjustbox}
\usepackage{doi}

\author{Shivanand Yadav}
\affiliation{Department of Physics, Bundelkhand University, Jhansi 284128, India}
\author{Raihan Ahammed}
\affiliation{Department of Materials Science \& Engineering, Indian Institute of Technology, Kanpur, Kanpur 208016, India}
\author{Umer Ali}
\affiliation{Department of Electrical Engineering, Indian Institute of Technology, Kanpur, Kanpur 208016, India}	
\author{B. S. Bhadoria}
\affiliation{Department of Physics, Bundelkhand University, Jhansi 284128, India}
\author{Yogesh S. Chauhan}
\affiliation{Department of Electrical Engineering, Indian Institute of Technology, Kanpur, Kanpur 208016, India}	
\author{Amit Agarwal}
\email{amitag@iitk.ac.in}
\affiliation{Department of Physics, Indian Institute of Technology, Kanpur, Kanpur 208016, India}
\author{Somnath Bhowmick}
\email{bsomnath@iitk.ac.in}
\affiliation{Department of Materials Science \& Engineering, Indian Institute of Technology, Kanpur, Kanpur 208016, India}

\title{Interface-Engineered Piezoelectricity and Electromechanical Coupling in Two-Dimensional van der Waals Heterostructures}

\begin{document}


\begin{abstract}
Interfaces between dissimilar two-dimensional (2D) materials provide a route to engineer electromechanical responses that are not intrinsic to the individual constituent layers. Here, we use high-throughput density functional theory (DFT) calculations to investigate 461 van der Waals heterostructures constructed from 57 2D monolayers and establish a materials-design framework linking work-function mismatch, interfacial charge redistribution, dipole formation, and strain-dependent electromechanical response. We find that heterostructure formation generates out-of-plane interfacial dipoles via charge redistribution between the constituent layers, while biaxial in-plane strain systematically modulates them. The resulting heterostructures exhibit two distinct regimes: metallic interfaces with strain-tunable interfacial dipoles and semiconducting heterostructures with well-defined piezoelectric responses. For the latter, Berry-phase (BP) calculations independently validate the charge-density-difference (CDD) approach, which is generally applicable to all representative systems, with piezoelectric coefficients calculated via two methods agreeing to within 10\%. Among 24 selected heterostructures, effective biaxial interfacial electromechanical coefficients reach 83.35 pCm\(^{-1}\). These results demonstrate that the interface can serve as the active element governing electromechanical coupling and provide a computationally guided strategy for designing strain-tunable out-of-plane electromechanical responses in 2D van der Waals heterostructures.
\end{abstract}


\section{Introduction}
The discovery of 2D materials has opened new opportunities in nanotechnology because of their remarkable structural, electronic, and mechanical properties. Their atomic-scale thickness, mechanical flexibility, and strong sensitivity to external perturbations make them promising for a wide range of nanoscale applications. Among these materials, graphene played a central role in establishing the field of 2D materials, while subsequent studies have revealed many other monolayers with semiconducting, magnetic, ferroelectric, and piezoelectric properties~\cite{10.1021/nn1029789, puri2018, PhysRevB.102.035420, naseer2024room, zelisko2014anomalous}. 

In recent years, considerable attention has been devoted to vertically stacking different monolayers to form van der Waals (vdW) heterostructures (HSs)~\cite{novoselov20162d}. Because weak vdW interactions hold the constituent layers together, they can be combined without the strict lattice-matching requirements of conventional HSs. At the same time, interfacial charge transfer, orbital hybridization, and symmetry breaking can generate properties that are absent in the isolated monolayers. As a result, vdW HSs have emerged as versatile platforms for electronic, optoelectronic, sensing, and energy-harvesting applications~\cite{yuan2020enhanced, chen2021multilayered, yadav2025high}. Several 2D HSs, including PtSe$_2$/PtTe$_2$~\cite{li2023van} and MoS$_2$/PtSe$_2$~\cite{rizu2025uv}, have already been investigated experimentally.

Among the many functionalities of 2D materials, electromechanical coupling or piezoelectricity is particularly attractive because it enables the direct conversion of mechanical deformation into electrical polarization~\cite{blonsky2015ab}. The piezoelectric effect refers to the generation of an electric response under applied strain or stress and is widely used in sensors, transistors, wearable electronics, and energy-harvesting devices~\cite{wang2006piezoelectric, ong2012engineered}. Piezoelectric materials are also being explored for harvesting mechanical energy from ambient sources such as human motion, wind, flowing water, and ocean waves, offering a potential route toward sustainable, self-powered technologies. Piezoelectricity requires the absence of inversion symmetry. Although many bulk crystals are centrosymmetric and therefore nonpiezoelectric, reducing a material to the monolayer limit can alter its symmetry and activate piezoelectric behavior. This possibility was first predicted theoretically and later confirmed experimentally in transition-metal dichalcogenides (TMDCs), including MoS$_2$ and WS$_2$~\cite{duerloo2012intrinsic, zhu2015observation}. These studies established 2D materials as promising candidates for nanoscale piezoelectric and piezotronic applications. 

Most conventional TMDC monolayers primarily exhibit in-plane piezoelectricity because they retain mirror symmetry along the out-of-plane direction. In contrast, out-of-plane piezoelectricity can emerge when this mirror symmetry is broken. Janus monolayers provide a natural route to achieving such structural asymmetry because different atomic species occupy the top and bottom surfaces of the layer. This asymmetry generates an intrinsic out-of-plane dipole and allows vertical polarization. Representative examples include MoSSe~\cite{dong2017large}, VSSe~\cite{zhang2019first}, and MoTO (T=S, Se, or Te)~\cite{dong2017large},  MoXSiN$_2$ (X = S, Se, Te)~\cite{gan2024emergence}, MX$_2$Y (M=Ti, Zr, Hf; X$\not=$Y=S, Se)~\cite{ahammed2020ultrahigh}, and BMX$_2$ (M = Ga, In; X = S, Se)~\cite{bezzerga2023ferro}.  

A similar symmetry-breaking mechanism can also arise in asymmetric vdW HSs composed of dissimilar monolayers. In such systems, differences in chemical composition, electronegativity, and interfacial bonding can induce charge redistribution and an out-of-plane dipole. Several HSs, including MoS$_2$/WSe$_2$~\cite{yu2018piezoelectricity}, GaN/BP with $d_{33} \approx 40\,\text{pmV}^{-1}$~\cite{mohanta2019superhigh}, and Na$_2$Tc/Cs$_2$S with $e_{33} \approx -59.68\,\text{pCm}^{-1}$~\cite{arora2023negative}, have been reported to exhibit strong out-of-plane piezoelectric responses. These findings demonstrate that interfacial engineering offers an effective strategy for controlling vertical polarization and enhancing piezoelectricity in 2D systems.

\begin{figure}
\centering{\includegraphics[width=0.9\linewidth]{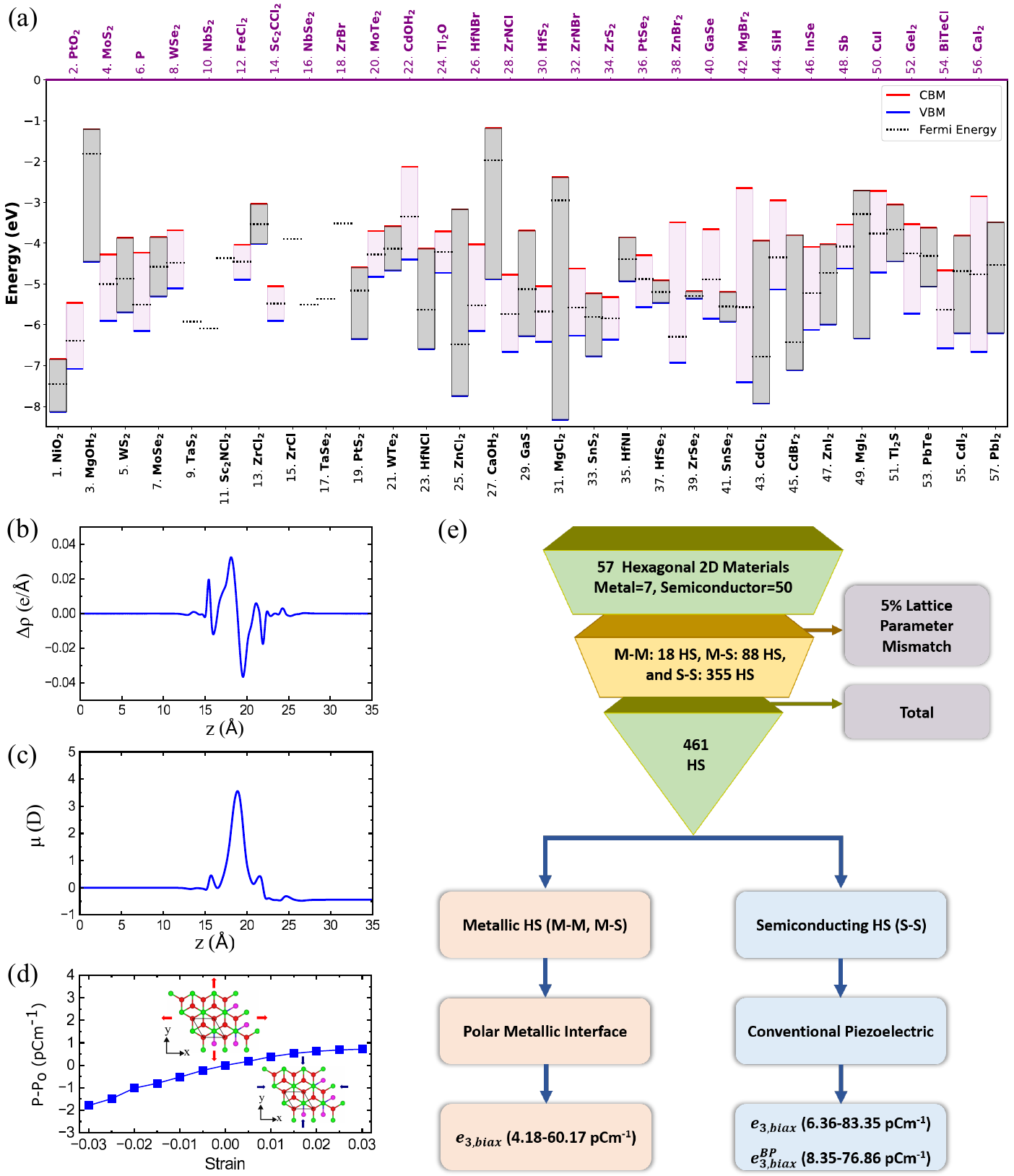}}
\caption{Workflow adopted in this work: (a) Calculated band edge positions [CBM: red, VBM: blue line], and Fermi energy [black dotted line] for the monolayer building blocks, referenced to the vacuum energy (0~eV). For metallic systems, only the Fermi energy is shown, indicating the absence of a band gap. The materials are numbered sequentially in order of increasing lattice parameter. (b) Work function difference provides the primary thermodynamic driving force for charge redistribution at the interface, (c) resulting in an out-of-plane dipole moment, (d) which can be engineered by in-plane strain. (e) High-throughput framework for classifying 461 heterostructures. The top 8 HSs in each category, ranked by polarization magnitude, are selected for electromechanical response studies, yielding coefficients ranging from 4.18 to 83.35 pCm$^{-1}$, comparable to that of known 2D materials.}
\label{fig1}
\end{figure}

Using high-throughput density functional theory (DFT) calculations, we systematically investigate 461 van der Waals heterostructures constructed from 57 two-dimensional monolayers to establish how interface formation can be used to engineer out-of-plane electromechanical responses. We consider metal–metal (M–M), metal–semiconductor (M–S), and semiconductor–semiconductor (S–S) heterostructures and examine the relationship between work-function mismatch, interfacial charge redistribution, dipole formation, and the response to biaxial in-plane strain. This analysis reveals two distinct electromechanical regimes. In M–M and M–S heterostructures, interface-induced charge redistribution produces strain-tunable out-of-plane interfacial dipoles, providing an electromechanical response without invoking a conventional bulk polarization in metallic systems. In S–S heterostructures, the interfacial dipole can be expressed as a macroscopic polarization and quantified within the conventional piezoelectric framework. We further validate the charge-density-difference approach against Berry-phase calculations for representative S–S heterostructures. The complete workflow is illustrated in Figure~\ref{fig1}. By identifying the interface, rather than the isolated monolayer, as the active element governing the electromechanical response, this work provides a materials-design strategy for developing strain-tunable out-of-plane electromechanical functionality in 2D van der Waals heterostructures. 

\section{Computational Methodology}
All first-principles calculations were performed within density functional theory using the plane-wave pseudopotential method as implemented in Quantum ESPRESSO~\cite{giannozzi2009quantum, PhysRevB.50.17953}, and VESTA was used for crystal structure visualization~\cite{momma2008vesta}. The exchange-correlation interaction was described using the Perdew–Burke–Ernzerh of generalized-gradient approximation~\cite{PhysRevLett.77.3865}, and van der Waals interactions were included using the Grimme DFT-D correction~\cite{Grimme2010}. Projector-augmented-wave pseudopotentials were employed with a plane-wave kinetic-energy cutoff of 50 Ry for wavefunctions and 500 Ry for charge density. The Brillouin zone was sampled using a $\Gamma$-centered  \(12\times12\times1\) Monkhorst–Pack \(k\)-point mesh. Electronic occupations were smeared using the Marzari–Vanderbilt (cold) smearing scheme with a broadening width of 0.01 Ry to aid self-consistent-field convergence. Structural relaxations were performed until the residual forces and total-energy changes were below \(10^{-3}\) Ry/a.u. and \(10^{-4}\) Ry, respectively.

A vacuum spacing of 35~\AA~ was introduced along the out-of-plane direction to minimize interactions between periodically repeated heterostructures. To prevent artificial electrostatic interactions between periodic images along the vacuum direction, a dipole correction was applied self-consistently, introducing a compensatory sawtooth potential in the vacuum region to ensure the correct electrostatic boundary conditions for the asymmetric slab geometry. The following calculations were performed using the DFT-predicted ground state charge density.

To quantify the electronic dipole moment, first, the planar-averaged charge-density difference (CDD) was calculated as,
\begin{equation}
\Delta \rho(z) = \int  \left( \rho_{\mathrm{HS}} - \rho_{\mathrm{M_1}} - \rho_{\mathrm{M_2}} \right) dx\,dy~,
\label{eq:3}
\end{equation}
where $\rho_{\mathrm{HS}}$ is the charge density of the heterostructure, while $\rho_{\mathrm{M_1}}$ and $\rho_{\mathrm{M_2}}$ are the charge densities of the isolated constituent monolayers calculated in the same atomic configurations as in the HS. The positive and negative values of $\Delta\rho(z)$ indicate charge accumulation and depletion, respectively. The CDD therefore provides direct insight into the interfacial charge redistribution and its influence on the electronic properties of the HS. The cumulative planar-averaged charge transfer, $\Delta Q(z)$, was obtained by integrating the CDD along the out-of-plane direction as,
\begin{equation}
\Delta Q(z) = \int_{-\infty}^{z} \Delta \rho(z')\,dz'~.
\label{eq:4}
\end{equation}
Similarly, the cumulative dipole profile was calculated as,
\begin{equation}
\mu(z) = \int_{-\infty}^{z} z'\,\Delta \rho(z')\,dz'~,
\label{eq:5}
\end{equation}
and the total interface induced dipole was obtained from the cumulative dipole profile as,
\begin{equation}
\mu_z=\mu(z_{max}),
\label{eq:5a}
\end{equation}
for a simulation cell spanning from $[0,z_{max}]$ in the out-of-plane direction. Because the charge-density difference integrates to zero over the simulation cell, the resulting total induced dipole is independent of the choice of origin along \(z\). The CDD-derived interfacial polarization was obtained as, 
\begin{equation}
    P_3=\frac{\mu_z}{A},
\end{equation}
where $A$ is the area of the simulation cell.

The piezoelectric stress-charge coefficient is conventionally defined as~\cite{zhang2019first}, $e_{ij} = \frac{\partial P_i}{\partial \epsilon_{j}}$ and $e_{ij} = d_{ik} C_{kj}$, where \(P_i\) denotes the polarization component, \(\epsilon_{j}\) is the strain tensor, $d_{ik}$ piezoelectric strain-charge coefficient and \(C_{kj}\) is the elastic stiffness constant. The electromechanical response was calculated by applying an in-plane biaxial strain and measuring the change in the out-of-plane dipole moment. The ``effective biaxial interfacial electromechanical stress coefficient'' was defined as,
\begin{equation}
e_{3,biax} = \frac{\partial P_3}{\partial \epsilon_{biax}}~.
\end{equation}
Biaxial strains ranging from \(-3\%\) to \(+3\%\) in increments of 0.5\% were considered. The effective biaxial coefficient was obtained from a linear least-squares fit of \(P_3\) versus $\epsilon_{biax}$. The applied biaxial strain was quantified by
\begin{equation}
\epsilon = \frac{a - a_0}{a_0}~,
\end{equation}
where \(a\) is the strained lattice parameters, and \(a_0\) is the corresponding equilibrium value. For each imposed biaxial strain, the in-plane lattice vectors were scaled according to \(a=b=(1+\epsilon)a_0\), while the vacuum dimension was kept fixed. The internal atomic coordinates were subsequently relaxed at fixed strained lattice vectors until the force convergence criterion was satisfied. The coefficient \(e_{3,\mathrm{biax}}\) defined here corresponds to the change in out-of-plane two-dimensional polarization under in-plane biaxial strain, \(e_{3,\mathrm{biax}}=\partial P_3/\partial\epsilon_{\mathrm{biax}}\). It should therefore not be directly compared numerically with the conventional \(e_{33}\), or \(e_{31}\) coefficients, which describe the response to uniaxial out-of-plane strain.

For semiconducting S–S heterostructures, the conventional total polarization and piezoelectric response were additionally evaluated using the Berry-phase formalism~\cite{king1993}, and it is denoted by $e^{BP}_{3,biax}$ to clearly differentiate from the CDD-based $e_{3,biax}$. This allows the electronic contribution obtained from the CDD analysis to be compared with the total polarization response obtained from the BP approach, which includes both electronic and ionic contributions. To get the 2D polarization values (pCm$^{-1}$), we multiplied the 3D polarization values obtained from DFT output by the physical thickness of the heterostructure, excluding the vacuum region. 

The Berry phase method does not apply to metallic systems because the modern theory of polarization based on the Berry phase is formulated for insulating systems with a set of occupied electronic states separated from the unoccupied states by a finite band gap~\cite{resta2002insulators, xu2020two}. In metallic systems, the Fermi level intersects one or more bands, resulting in partially occupied electronic states and preventing an unambiguous definition of the bulk polarization through the conventional Berry-phase methods. Instead, CDD provides an alternative method for metallic systems (M--M and M--S) by capturing the strain-induced redistribution of electronic charge and the corresponding change in the out-of-plane dipole moment, without requiring an insulating band structure.

\section{Results and discussion}
\paragraph{High-Throughput Computational Screening:} We developed a high-throughput screening workflow, starting with 57 two-dimensional (2D) monolayers \cite{mounet2018two}, and constructed 461 heterostructures (HSs) with lattice mismatches of less than 5\%.  The band alignment of the examined 2D materials is illustrated in Figure~\ref{fig1}(a), showing a thorough comparison of their electronic structures. The electronic energies in Figure~\ref{fig1}(a) are referenced to the vacuum level, set to \(0\) eV. As a result, the Fermi energies are negative. The work function is defined as
\begin{equation}
\Phi=E_{\rm vac}-E_F,
\label{eqwf}
\end{equation}
which is therefore positive for all systems considered here.

Due to the variation in work function among the examined monolayers, the formation of a van der Waals HS leads to spontaneous electron transfer from the material with the lower work function to that with the higher work function until their Fermi levels are aligned. This charge redistribution gives rise to an intrinsic built-in electric field across the interface, which influences the electronic properties of the HS; such as interfacial charge redistribution, resulting in a dipole moment in the out-of-plane direction, and how it changes with strain, were subsequently calculated [Figure~\ref{fig1}(b)-(d)].
 
These 461 HSs include 18 metal–metal (M–M), 88 metal–semiconductor (M–S), and 355 semiconductor–semiconductor (S–S) heterostructures, as shown in Figure~\ref{fig1}(e). For each HS, six different stacking configurations were considered, corresponding to rotational angles of \( 0^\circ \), \( 60^\circ \), \( 120^\circ \), \( 180^\circ \), \( 240^\circ \), and \( 300^\circ \). The energetically most favorable configuration was then selected based on the minimum total energy among these stacking arrangements. We subsequently selected 8 HSs from each category with higher out-of-plane polarization within their respective categories, yielding a total of 24 candidate materials. These materials were then investigated in detail to quantify their electromechanical response.

\begin{figure}
\centering{\includegraphics[width=0.8\linewidth]{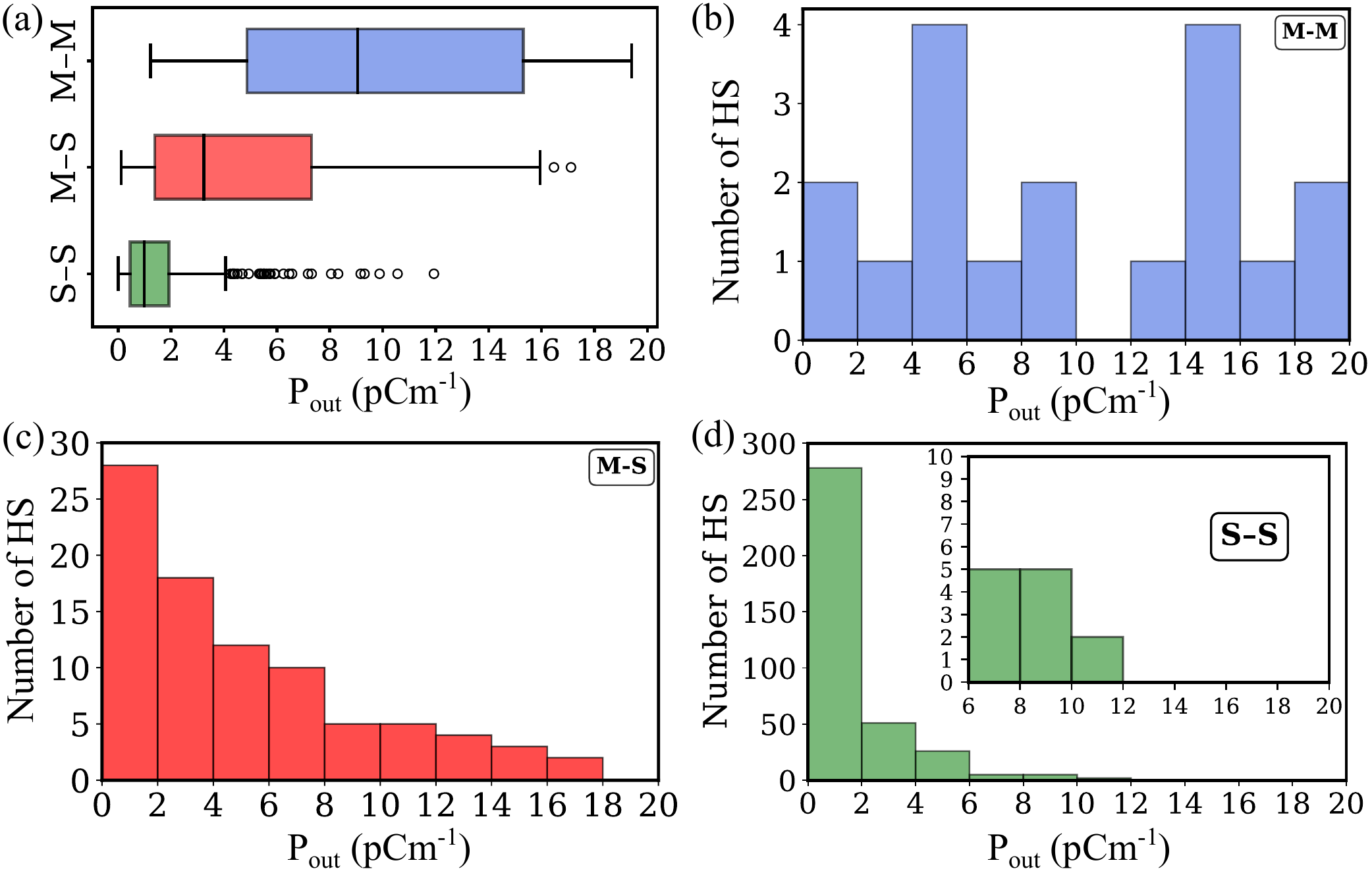}}
\caption{(a) Polarization distribution for S-S (green), M-S (red), and M-M (blue) HSs. In the box-and-whisker plot, the box represents the inter-quartile range (from the first quartile, $Q1~ (25\%)$, to the third quartile, $Q3~ (75\%)$), and the median (50\%) is indicated by a line inside the box. Outliers are depicted as circles. A comparative analysis of polarization behavior is further presented for different types of HSs: (b) M-M, (c) M-S, and (d) S-S. }
\label{fig2}
\end{figure}

 The distribution of out-of-plane polarization for different HS types is shown in Figure~\ref{fig2}(a), with M–M, M–S, and S–S systems illustrated separately. The box-and-whisker plot shows that M–M HSs exhibit the highest out-of-plane polarization, with quartile values of approximately 4.70, 9.63, and 15.62~pCm$^{-1}$, along with a relatively broad distribution. In comparison, M–S HSs show intermediate behavior, with quartile values of about 1.37, 3.24, and 7.42~pCm$^{-1}$, and a moderate spread. By contrast, S–S HSs exhibit the lowest polarization values, with quartiles of about 0.46, 0.99, and 2.04~pCm$^{-1}$, together with a much narrower distribution. This trend can be attributed to enhanced interfacial charge transfer and stronger symmetry breaking in M–M HSs, which promote larger out-of-plane polarization. In contrast, the reduced charge redistribution and weaker interfacial effects in S–S HSs lead to smaller polarization values.

The polarization distribution of M–M HSs within the range of $1.22-19.41$~pCm$^{-1}$ is shown in Figure~\ref{fig2}(b), while the corresponding distribution for M–S HSs, spanning $0.11-17.11$~pCm$^{-1}$, is presented in Figure~\ref{fig2}(c). Interestingly, finite out-of-plane polarity is obtained even for metallic heterostructures. Here, inversion-symmetry breaking and interfacial charge redistribution can generate a nonzero dipole despite the presence of mobile carriers. Figure~\ref{fig2}(d) shows the distribution for S–S HSs, which covers the range of $0.02-11.93$~pCm$^{-1}$. Most S–S HSs are concentrated in the low-polarization range of $0.02-2$~pCm$^{-1}$, whereas the inset highlights a smaller number of systems with higher polarization values $>6$~pCm$^{-1}$. Several high polarization HSs found here are comparable to other two-dimensional polar materials, such as 2D bimetal phosphates ($P_{\mathrm{out}} = 1$--10~pC\,m$^{-1}$)~\cite{ma2021large}, Co$_2$Se$_3$ ($P_{\mathrm{out}} = 4$~pC\,m$^{-1}$)~\cite{tang2021first}, semiconducting 2D MXenes ($P_{\mathrm{out}} = 3$--17~pC\,m$^{-1}$)~\cite{zhang2020first}, and group III--V monolayers ($P_{\mathrm{out}} = 9.48$--13.96~pC\,m$^{-1}$)~\cite{naseer2024room}.

Next, we selected 8 HSs from each category, ranked by polarization magnitude, and investigated their electromechanical response, yielding $e_{3,biax}$ coefficients ranging from 4.18 to 83.35 pCm$^{-1}$. Interestingly, the magnitude of the static interfacial polarization does not by itself determine the strength of the electromechanical response. Specifically, few S--S HSs are found with higher $e_{3,biax}$ values, despite having less $P_{\mathrm{out}}$ compared to M--M HSs. Based on the calculated $e_{3,biax}$ values, we selected a few HSs with very high $e_{3,biax}$ for further analysis. Specifically, ZrCl–NbSe$_2$, TaS$_2$–ZrCl, and NbS$_2$–ZrCl were selected from the M–M category; ZrBr–ZrS$_2$, ZrBr–HfS$_2$, and ZrBr–ZrNCl were selected from the M–S category; and Cd(OH)$_2$–ZrS$_2$, Sc$_2$CCl$_2$–Cd(OH)$_2$, and Tl$_2$O-SnS$_2$ were selected from the S–S category. The following sections detail the electronic band structure and electromechanical response of selected HSs, enabling consistent comparisons across different families and uncovering the underlying physics.

\begin{figure*}
\centering{\includegraphics[width=1\linewidth]{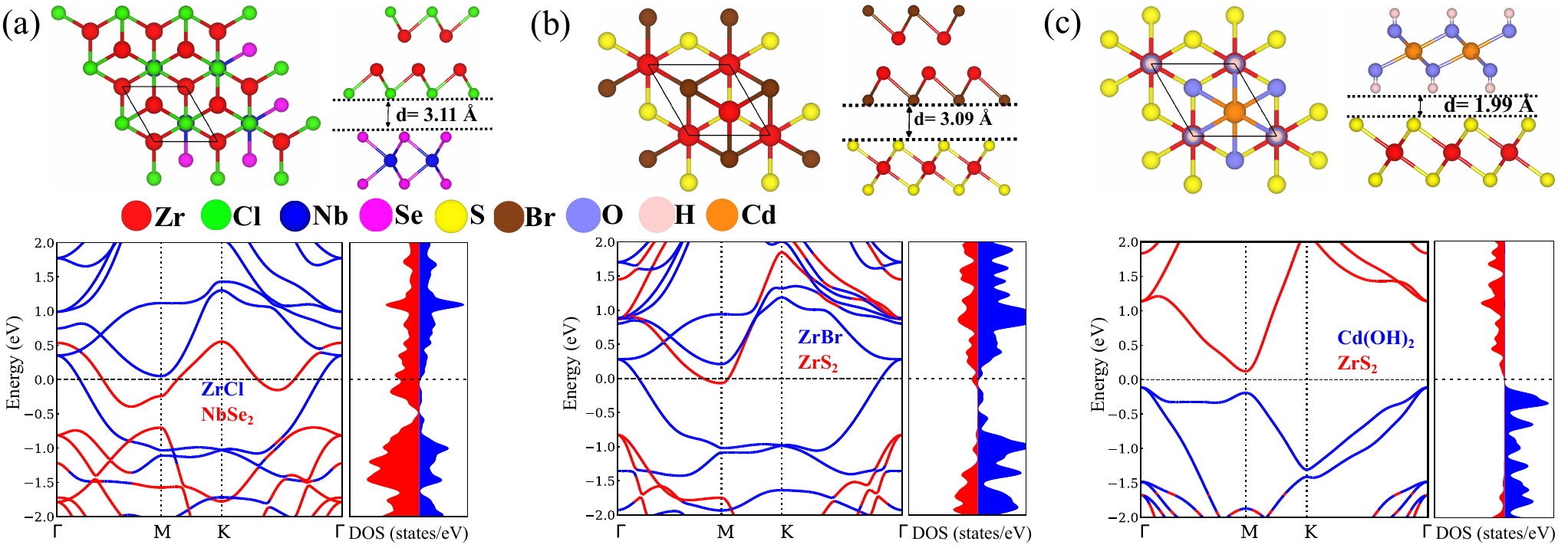}}
\caption{Top and side views, along with the layer-resolved electronic band structures and layer-projected density of states (DOS) of (a) ZrCl–NbSe$_2$ M--M HS, (b) ZrBr–ZrS$_2$ M--S HS, (c) Cd(OH)$_2$–ZrS$_2$ S-S HS. The parameter $d$ denotes the interlayer distance between the constituent monolayers.}
\label{fig3}
\end{figure*}

\paragraph{Crystal Geometry and Stability of Heterostructures:} Before investigating the selected HSs, it is important to evaluate the properties of the constituent monolayers, particularly their crystal structures, because lattice mismatch serves as the first filter during heterostructure formation. The electronic band structure is also important for classifying the HSs into the M-M, M-S, and S-S categories. The electronic band structures show that ZrCl, NbSe$_2$, TaS$_2$, NbS$_2$, ZrBr, Sc$_2$NCl$_2$, and TaSe$_2$ are metallic. In contrast, Cd(OH)$_2$, Tl$_2$O, and FeCl$_2$ are direct-band-gap semiconductors with band gaps of 2.36, 1.02, and 0.86~eV, whereas ZrS$_2$, HfS$_2$, ZrNCl, Sc$_2$CCl$_2$, SnS$_2$, PtS$_2$, and ZrCl$_2$ are indirect-band-gap semiconductors with band gaps of 1.04, 1.36, 1.89, 0.85, 1.62, 1.76, and 1.03~eV, respectively. These results, along with optimized lattice parameters, are summarized in Figure~S1, Supporting Information (SI).

During heterostructure formation, the lattice mismatch was kept below 5\% in all cases. The optimized crystal structure of the representative M--M HS, ZrCl-NbSe$_2$, is shown in Figure~\ref{fig3}(a). It has a lattice constant of $a = 3.44$~\text{\AA} and an interlayer distance of $d = 3.11$~\text{\AA}. The remaining M--M HSs, TaS$_2$-ZrCl and NbS$_2$-ZrCl, are provided in Figures~S2(a-b), Supporting Information. 

For the M--S HSs, the optimized structure of ZrBr-ZrS$_2$ is shown in Figure~\ref{fig3}(b), with a lattice constant of $a = 3.55$~\text{\AA} and an interlayer distance of $d = 3.09$~\text{\AA}. The other two M--S HSs, ZrBr-HfS$_2$ and ZrBr-ZrNCl, are shown in Figures~S2(c-d), Supporting Information. 

Similarly, the optimized structure of the representative S--S HS, Cd(OH)$_2$-ZrS$_2$, is shown in Figure~\ref{fig3}(c). It has a lattice constant of $a = 3.58$~\text{\AA} and an interlayer distance of $d = 1.99$~\text{\AA}. The remaining S--S HSs, Sc$_2$CCl$_2$-Cd(OH)$_2$ and Tl$_2$O-SnS$_2$ are presented in Figures~S2(e-f), Supporting Information.

To evaluate the energetic stability of the HSs, we calculated the binding energy, E$_b$, using the following expression:
\begin{equation}
E_b = \frac{E_{HS} - E_{m_1} - E_{m_2}}{A}~,
\end{equation}
where E$_{HS}$ is the total energy of the HS, E$_{m_1}$ and E$_{m_2}$ are the total energies of the isolated constituent monolayers, and A is the area of the optimized HS. A negative value of E$_b$ indicates that the formation of the HS is energetically favorable.

For the M--M HSs, the binding energies of ZrCl-NbSe$_2$, TaS$_2$-ZrCl, and NbS$_2$-ZrCl are $-19.22$, $-22.69$, and $-18.66$~meV/\text{\AA}$^{2}$, respectively. For the M--S HSs, ZrBr-ZrS$_2$, ZrBr-HfS$_2$, and ZrBr-ZrNCl have binding energies of $-16.96$, $-21.76$, and $-14.13$~meV/\text{\AA}$^{2}$, respectively. Similarly, the S--S HSs Cd(OH)$_2$-ZrS$_2$, Sc$_2$CCl$_2$-Cd(OH)$_2$, and Tl$_2$O-SnS$_2$ exhibit binding energies of $-14.86$, $-11.61$, and $-30.25$~meV/\text{\AA}$^{2}$, respectively. These binding energy values are comparable to other stable van der Waals HSs reported in the literature, such as GaAs/HfS$_2$ ($-11.61$~meV/\text{\AA}$^{2}$)~\cite{zhang2025promising}, XY$_2$/Sc$_2$CCl$_2$ (X = Nb, Ni, Ti, V, Mn, Ta; Y = S, Se) ($-13.37$~meV/\text{\AA}$^{2}$)~\cite{xie2026modulating}, 1T-TaS$_2$/black phosphorus ($-20.70$~meV/\text{\AA}$^{2}$)~\cite{wang2019modulating}, and NbS$_2$/MoSe$_2$ ($-30.64$~meV/\text{\AA}$^{2}$)~\cite{khang2024achieving}.

We further calculated the phonon dispersion of all nine heterostructures (3 M--M, 3 M--S, 3 S--S), which were earlier selected for detailed electronic and electromechanical investigation. The corresponding phonon dispersion spectra are provided in Figure~S3, Supporting Information, showing the absence of imaginary phonon frequencies throughout the Brillouin zone. Thus, the negative binding energies indicate favorable interlayer binding, while the absence of imaginary phonon modes confirms the dynamical stability of the investigated heterostructures. 

\paragraph{Electronic Band structure:} After evaluating the stability, we then analyzed the electronic band structures of the constructed HSs in detail. The electronic band structures of the investigated HSs provide important insight into their electronic properties. As shown in Figure~\ref{fig3}(a), M--M heterostructure ZrCl-NbSe$_2$ is metallic with predominantly NbSe$_2$-derived states around E$_F$. Comparing the work function values of the constituent monolayers (Figure~\ref{fig1}(a)), the lower work function of ZrCl relative to NbSe$_2$ drives interfacial electron transfer from ZrCl to NbSe$_2$ for Fermi-level equilibration. Since NbSe$_2$ is metallic, there are available states into which the transferred charge can redistribute. This charge transfer is expected to induce an interfacial dipole, which will be discussed separately later using the charge-density difference and Bader charge analysis.    

The M--S heterostructure ZrBr-ZrS$_2$ (Figure~\ref{fig3}(b)) also shows overall metallic character. The layer-projected DOS shows a fairly pronounced energy separation between the contributions of the two layers. In particular, the states on the occupied side just below the Fermi energy have a stronger ZrS$_2$ contribution. In contrast, the low-lying unoccupied states just above the Fermi energy have a stronger ZrBr contribution. The calculated work functions show that ZrBr has a lower work function than ZrS$_2$ (Figure~\ref{fig1}(a)), suggesting that electrons are preferentially transferred from occupied states of metallic ZrBr to unoccupied CBM states of ZrS$_2$ upon interface formation until Fermi-level equilibration is achieved.

The S--S heterostructure Cd(OH)$_2$-ZrS$_2$ (Figure~\ref{fig3}(c)) has an indirect band gap of 0.50~eV. Rather than both layers contributing equally to both the valence and conduction regions, the ZrS$_2$-derived states and Cd(OH)$_2$-derived states are energetically separated.  In particular, the conduction band minimum (CBM) and valence band maximum (VBM) states are contributed by ZrS$_2$ and Cd(OH)$_2$, respectively. This spatial separation of band-edge states is characteristic of type-II band alignment, which is beneficial for photocatalytic and photovoltaic applications due to efficient electron-hole separation. Because Cd(OH)\(_2\) has a lower work function than ZrS\(_2\) (Figure~\ref{fig1}(a)), electrons transfer from Cd(OH)\(_2\) to ZrS\(_2\) upon heterostructure formation to achieve Fermi-level equilibration. The type-II alignment further spatially separates the band-edge states, with the CBM and VBM predominantly localized on ZrS\(_2\) and Cd(OH)\(_2\), respectively.

A similar trend is observed for other representative heterostructures, like TaS$_2$-ZrCl, NbS$_2$-ZrCl (M--M, Figure~S2(a)-(b), Supporting Information), ZrBr-HfS$_2$, ZrBr-ZrNCl (M--S, Figure~S2(c)-(d), Supporting Information), Sc$_2$CCl$_2$-Cd(OH)$_2$, and Tl$_2$O-SnS$_2$ (S--S, Figure~S2(e-f), Supporting Information). In type-II heterostructures, the spatial separation of the band-edge states favors interlayer charge redistribution, leading to pronounced electron accumulation and depletion on opposite sides of the interface. This enhanced charge separation increases the interfacial dipole moment and consequently the out-of-plane polarization. 

Note that, in this work, we are mainly interested in ground-state charge transfer, which should be distinguished from photoinduced electron–hole separation. While the former is driven by Fermi-level equilibration and work-function difference, the latter mainly depends on the electrostatic potential between the monolayers. The strain-dependent electrostatic potential profiles (Figures S4–S6, Supporting Information) further demonstrate the evolution of the interfacial potential step accompanying charge redistribution. The rest of the paper will focus on ground-state charge transfer driven by Fermi-level equilibration, related dipole moment, and its electromechanical response. 

\paragraph{Electromechanical Response of Heterostructures:} After understanding the driving force for charge redistribution in terms of work function and electronic band structure, let us now quantify the interface-induced electronic polarization and its response to mechanical strain. In the following text, the microscopic origin of the electromechanical coupling is elucidated by examining the strain-dependent evolution of the charge-density difference CDD [Eq.~\ref{eq:3}], integrated charge transfer $\Delta Q(z)$ [Eq.~\ref{eq:4}], cumulative dipole moment $\mu(z)$ [Eq.~\ref{eq:5}], and polarization under biaxial strain. The CDD provides direct information on electron accumulation and depletion at the interface, while the integrated charge transfer quantifies the charge transferred between the constituent layers. The resulting cumulative dipole moment reflects the spatial separation of positive and negative charge centers across the interface. Since polarization ($P$) originates from charge separation, strain-induced variations in charge transfer directly influence the polarization response. Therefore, the observed evolution of CDD, $\Delta Q$, dipole moment, and polarization provides direct evidence of electromechanical coupling in these van der Waals heterostructures.

\begin{figure}
\centering{\includegraphics[width=0.6\linewidth]{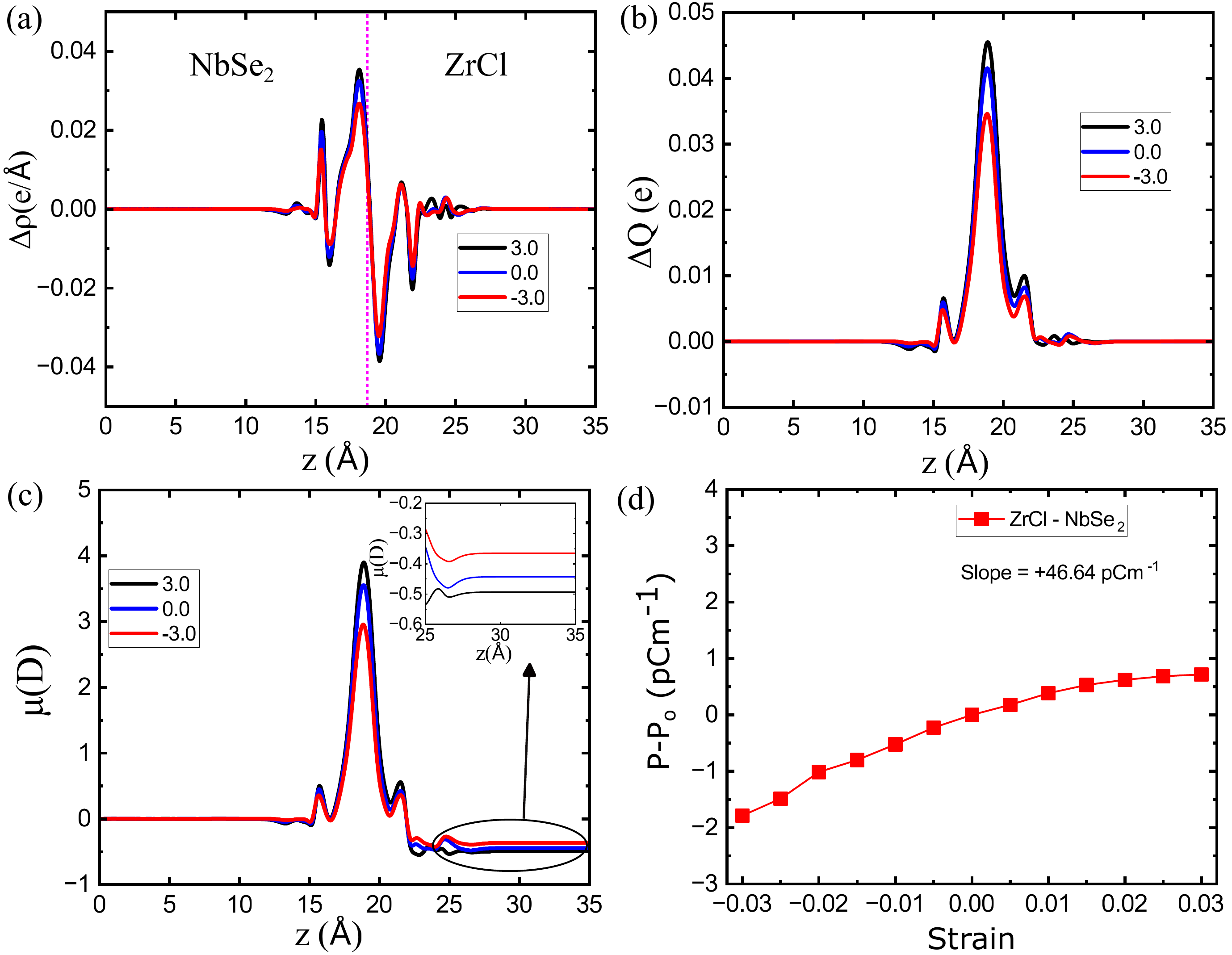}}
\caption{M–M HS ZrCl–NbSe$_2$: (a) Plane-averaged charge density difference (CDD), illustrating charge redistribution at the interface between the constituent monolayers. The dashed line indicates the interface. (b) Plane-averaged cumulative charge transfer ($\Delta Q$), depicting the interlayer charge transfer profile along the plane normal. (c) Interfacial dipole moment of the HS, with the inset showing its evolution under applied biaxial strain of $3\%, ~0\%,~  \text{and } -3\%$. (d) Variation of the out-of-plane polarization with strain.}
    \label{fig4}
\end{figure}

For the illustrative M--M HS ZrCl-NbSe$_2$, the CDD shows pronounced accumulation in the NbSe$_2$ layer and electron depletion in the ZrCl layer, as a result of spontaneous interfacial charge transfer from ZrCl to NbSe$_2$ (Figure~\ref{fig4}(a)). Bader charge analysis further confirms a net charge transfer of approximately $|0.04e|$ [Table S1, Supporting Information], which leads to the formation of an intrinsic interfacial dipole. Under biaxial strain, the distribution of accumulated and depleted charges evolves systematically, demonstrating that mechanical deformation modifies the interfacial charge distribution. This behavior is quantitatively reflected in the integrated charge-transfer curves shown in Figure~\ref{fig4}(b), where the transferred charge ($\Delta Q$) exhibits a clear strain dependence. The strain-induced variation in $\Delta Q$ directly affects the dipole moment, as shown in Figure~\ref {fig4}(c), since the dipole originates from the spatial separation of the charge accumulation and depletion regions. The inset of Figure~\ref{fig4}(c) reveals an increase in the magnitude of the dipole moment under tensile strain, indicating an enhanced interfacial charge separation. As shown in Figure~\ref{fig4}(d), polarization varies almost linearly with strain, resulting in a positive coefficient $e_{3,\mathrm{biax}}$ of 46.64~pCm$^{-1}$. The positive sign indicates that tensile strain promotes charge separation and enhances interfacial polarization.

A similar behavior is observed for the rest of the M--M HSs, TaS$_2$-ZrCl (Figure~S7(a--d), Supporting Information) and NbS$_2$-ZrCl (Figure~S8(a--d), Supporting Information); the evolution of CDD, $\Delta Q$, dipole moment, and polarization follows the same trend. The corresponding coefficients $e_{3,\mathrm{biax}}$ are 43.45 and 39.90~pCm$^{-1}$, respectively, further confirming that the electromechanical coupling in M--M heterostructures originates from strain-modulated interfacial charge transfer.

\begin{figure}
\centering{\includegraphics[width=0.6\linewidth]{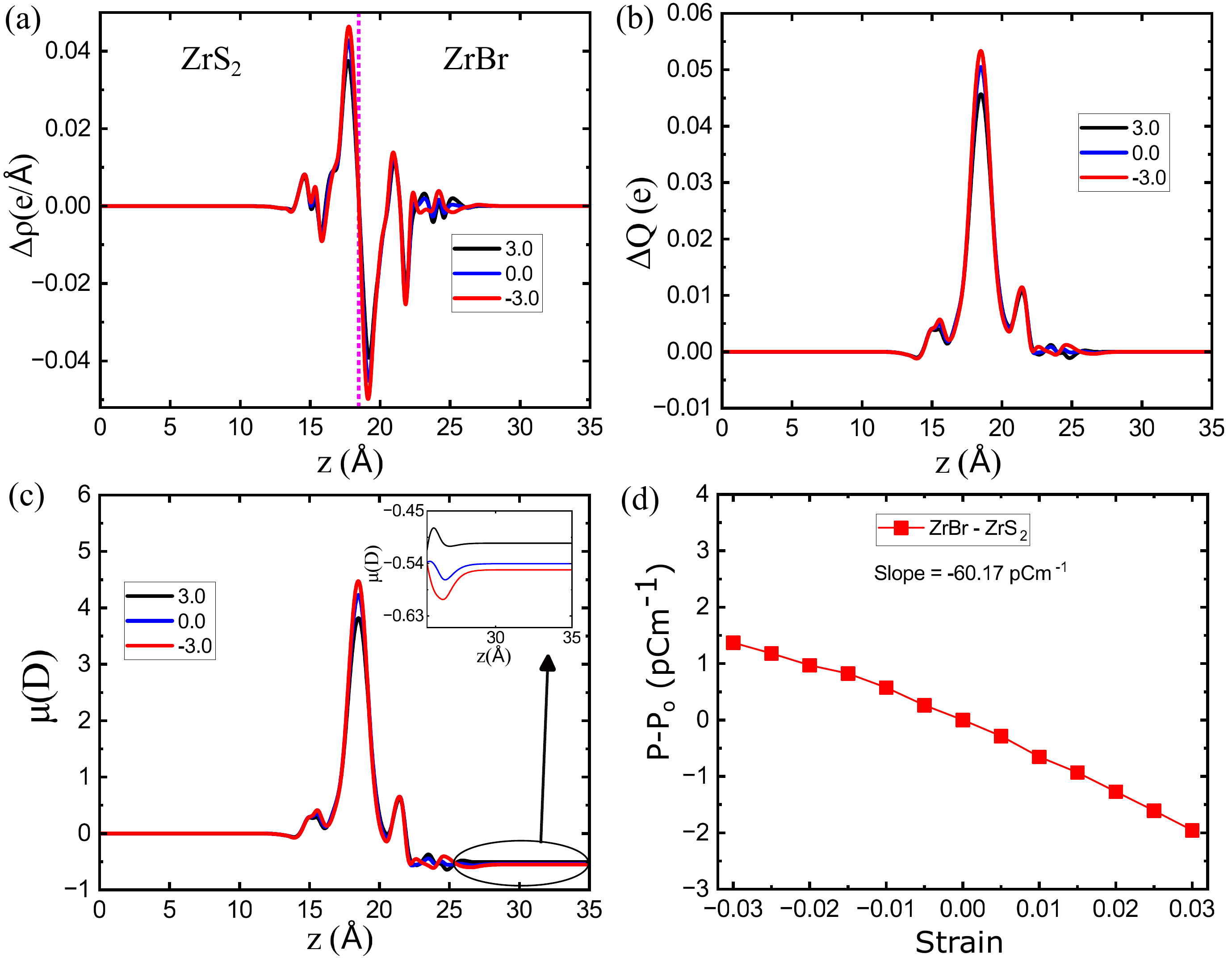}}
\caption{M–S HS ZrBr–ZrS$_2$: (a) Plane-averaged CDD. (b) Plane-averaged charge transfer ($\Delta Q$). (c) Interfacial dipole moment of the HS, with the inset showing its variation under biaxial strain of $3.0\%, ~0\%,~  \text{and } -3.0\%$. (d) Variation of out-of-plane polarization as a function of applied strain.}
\label{fig4a}
\end{figure}

For the illustrative M--S HS ZrBr-ZrS$_2$, the CDD shows pronounced electron accumulation in the ZrS$_2$ layer and electron depletion in the ZrBr layer, as a result of spontaneous interfacial charge transfer from ZrBr to ZrS$_2$ (Figure~\ref{fig4a}(a)). Bader charge analysis further reveals a net charge transfer of approximately $|0.06e|$ [Table S1, Supporting Information], leading to the formation of an intrinsic interfacial dipole. The integrated charge-transfer $\Delta Q$ exhibits pronounced strain-dependent variations (Figure~\ref{fig4a}(b)), as biaxial strain effectively modulates interfacial electron redistribution. As a consequence, the dipole moment magnitude decreases under tensile strain (Figure~\ref{fig4a}(c)). The polarization decreases almost linearly with increasing tensile strain, yielding a negative coefficient $e_{3,\mathrm{biax}}$ of $-60.17$~pCm$^{-1}$ (Figure~\ref{fig4a}(d)). The negative sign indicates that the tensile strain weakens the interfacial charge separation and reduces the effective dipole moment.

Among the rest of the M--S HSs, a similar trend is observed for ZrBr-HfS$_2$ (Figure~S9(a-d), Supporting Information), with a negative $e_{3,\mathrm{biax}}$ coefficient of $-58.91$~pCm$^{-1}$. In contrast, the ZrBr-ZrNCl heterostructure (Figure~S10(a-d), Supporting Information) shows an increase in the magnitude of the dipole moment and a corresponding enhancement of polarization under tensile strain, giving rise to a positive $e_{3,\mathrm{biax}}$ of 49.56~pCm$^{-1}$. The coexistence of positive and negative electromechanical responses within the M--S category demonstrates that the sign and magnitude of electromechanical coupling are governed primarily by strain-induced modifications to atomic bonding and band-alignment, which affect interfacial charge-transfer pathways, rather than by the overall heterostructure classification alone.

These results establish that metallic heterostructures can exhibit sizable, strain-tunable interfacial dipoles despite the absence of a conventional bulk polarization. We therefore refer to the M--M and M--S systems as polar metallic interfaces rather than piezoelectric materials. The physical situation changes for the S--S HSs, where the electronic structure permits a well-defined macroscopic polarization. For these systems, we evaluated the strain-induced polarization using both the CDD and Berry-phase approaches, thereby separating the interface-driven electronic contribution from the total electromechanical response.

\begin{figure}[h]
\centering{\includegraphics[width=0.6\linewidth]{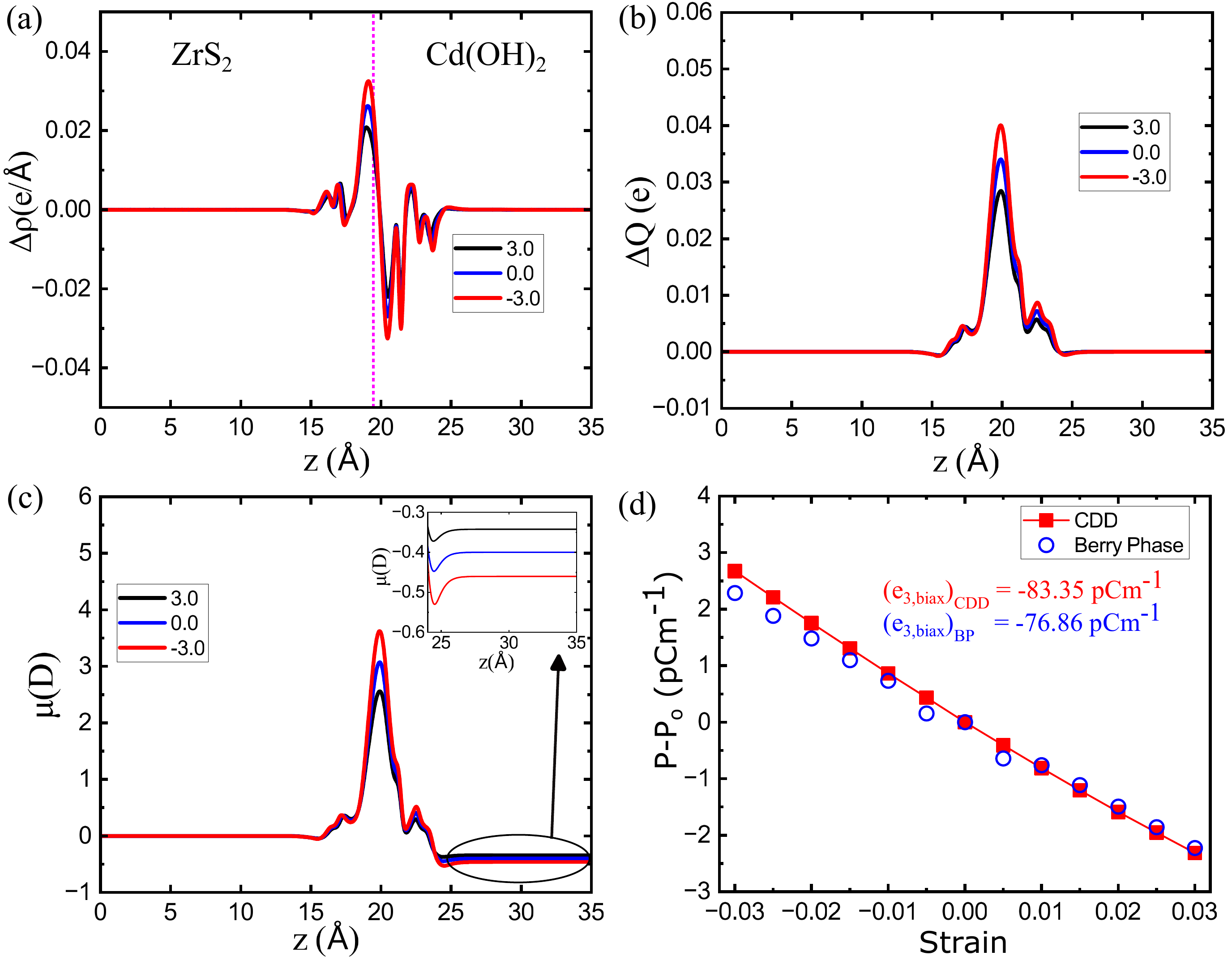}}
\caption{S–S HS Cd(OH)$_2$–ZrS$_2$: (a) Plane-averaged CDD. (b) Plane-averaged charge transfer ($\Delta Q$). (c) Interfacial dipole moment of the HS, with the inset illustrating its variation under biaxial strain of $3.0\%, ~0\%,~  \text{and } -3.0\%$. (d) Variation of out-of-plane polarization as a function of biaxial strain obtained using the CDD (red) and Berry phase (blue) methods.}
\label{fig4b}
\end{figure}

For the illustrative S--S HS Cd(OH)$_2$--ZrS$_2$, the CDD shows pronounced accumulation in the ZrS$_2$ layer and electron depletion in the Cd(OH)$_2$ layer (Figure~\ref{fig4b}(a)). Bader charge analysis further reveals a net charge transfer of $|0.06e|$ from Cd(OH)$_2$ to the ZrS$_2$ layer [Table S1, Supporting Information], leading to the formation of an intrinsic interfacial dipole. The integrated charge-transfer $\Delta Q$ has a strong strain-dependence (Figure~\ref{fig4b}(b)), as biaxial strain effectively modulates the interfacial electron redistribution, and the dipole moment magnitude decreases under tensile strain (Figure~\ref{fig4b}(c)). As a result, the polarization decreases linearly with increasing tensile strain, yielding a negative coefficient $e_{3,\mathrm{biax}}=-83.35$~pCm$^{-1}$, obtained from CDD. The Berry phase-derived coefficient was also calculated using the modern theory of polarization, yielding a value of $e^{BP}_{3,\mathrm{biax}}=-76.86$ pCm$^{-1}$ (Figure~\ref{fig4b}(d)). The relatively small difference ($\sim 10\%$) between the two approaches indicates that the electromechanical response in Cd(OH)$_2$--ZrS$_2$ is predominantly governed by the electronic contribution arising from strain-induced charge redistribution. The negative coefficient indicates that tensile strain weakens interfacial charge separation, thereby reducing the effective dipole moment.

A similar behavior is observed for the other S--S HSs, Sc$_2$CCl$_2$--Cd(OH)$_2$, which has large negative coefficients of $e_{3,biax}=-70.75$ pCm$^{-1}$ and $e^{BP}_{3,biax}=-65.15$ pCm$^{-1}$, obtained using the CDD and BP methods, respectively [Figure~S11(a--d), Supporting Information]. In contrast, Tl$_2$O--SnS$_2$ has positive coefficients of $e_{3,biax}=+69.53$ pCm$^{-1}$ and $e^{BP}_{3,biax}=+67.01$ pCm$^{-1}$, obtained using the CDD and BP methods, respectively [Figure~S12(a--d), Supporting Information]. The positive sign indicates that tensile strain promotes charge separation and enhances the interfacial polarization.

\begin{figure*}
\centering{\includegraphics[width=1\linewidth]{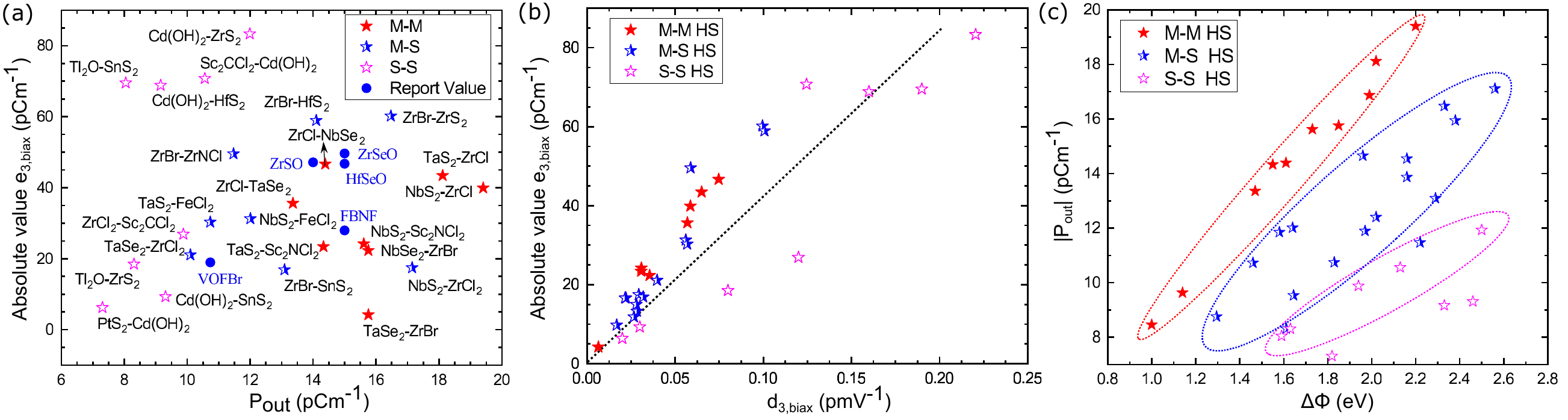}}
\caption{(a) Absolute value of the electromechanical coefficients: Stars represent $e_{3,biax}$ from the present study; circles represent literature \(e_{33}\) values for VOFBr~\cite{mahajan2023magnetoelectric}, ZrSO, ZrSeO, HfSeO, FBNF~\cite{pham2021enhanced}. The latter are added for qualitative context only; direct numerical comparison with \(e_{3,\mathrm{biax}}\) is not appropriate because of differences in strain geometry. (b) Linear correlation between $e_{3,biax}$ and the strain coefficient $d_{3,biax}$. (c) Variation of polarization with the work-function difference between the constituent monolayers, highlighting the role of work-function mismatch in governing interfacial charge redistribution and polarization.}
\label{fig5}
\end{figure*}

Analyzing the results presented above (see Figure~\ref{fig5}(a)), it is clear that M--M HSs generally have the largest out-of-plane polarization but moderate electromechanical coupling. On the other hand, the strongest electromechanical coupling is observed in the S--S HSs, although they have moderate out-of-plane polarization. The larger magnitude of $e_{3,biax}$ coefficient indicates a highly efficient conversion of mechanical deformation into electrical polarization. The enhanced electromechanical response of S--S heterostructures arises from their weak electrostatic screening, which enables strain-induced perturbations to be highly effective, such that even small changes in the lattice parameters can significantly modify the band-edge positions, charge-transfer characteristics, and dipole moment.

For the 24 selected HSs, the $e_{3,biax}$ values are listed in Table~\ref{tab1}. Overall, the values lie in the ranges of 4.18--46.64~pCm$^{-1}$ for M--M HSs, 16.86--60.17~pCm$^{-1}$ for M--S HSs, and 6.36--83.35~pCm$^{-1}$ for S--S HSs. These values are compared with reported $e_{33}$ or $e_{31}$ values for other 2D materials, including heterostructures and Janus materials, ranging from 1.0 to 96.88~pCm$^{-1}$.  Clearly, the calculated coefficients span a range comparable in magnitude to values reported for various 2D piezoelectric systems; however, these literature values are included only for qualitative context, as the strain geometry and dimensional normalization are not generally equivalent. Note that, although the metallic (M--M and M--S) HSs possess a static interface dipole, and they have a strain dependence or electromechanical response, it would be inappropriate to label them as promising piezoelectric material. Metallic systems are better classified as polar metallic interfaces with strain-tunable dipoles. On the other hand, the semiconducting S--S heterostructures provide the appropriate insulating/semiconducting electronic structure for a conventional Berry-phase-defined piezoelectric response. 

\begin{table}
\centering
\caption{Comparative analysis of the calculated out-of-plane polarization and $e_{3,biax}$ coefficients (pCm$^{-1}$). Overall, the values lie in the ranges of 46.64--4.18~pCm$^{-1}$ for M-M HSs, 60.17--16.86~pCm$^{-1}$ for M-S HSs, and 83.35--6.36~pCm$^{-1}$ for S-S HSs. Numbers in parentheses are $e^{BP}_{3,biax}$ values for S--S HSs. Literature-reported piezoelectric coefficients of 2D materials, including heterostructures and Janus materials, range from 1.0 to 96.88~pCm$^{-1}$. These are added for qualitative context only; as direct comparison with \(e_{3,\mathrm{biax}}\) is not appropriate because of differences in strain geometry.}
\resizebox{1.0\columnwidth}{!}{%
\begin{tabular}{|c c| c c|}
\hline
Present work & $|e_{3,biax}|(|e^{BP}_{3,biax}|)$ & Reported Material & $|e_{33}|, |e_{31}|$\\
\hline
ZrCl-NbSe$_2$  & 46.64            & Penta-graphene & 96.88 \cite{jia2018piezoelectric} \\
TaS$_2$-ZrCl  & 43.45             & VSSe & 76.50 \cite{yang2019coexistence} \\
NbS$_2$-ZrCl  & 39.90             & Pd$_4$S$_3$Te$_3$ & 74.33 \cite{li2024intrinsically} \\
ZrCl-TaSe$_2$  & 35.64            & MoSi$_2$N$_2$P$_2$ & 73.1 \cite{yin2023monolayer} \\
NbS$_2$-Sc$_2$NCl$_2$ & 24.20     & Y$_2$Se$_2$BrF & 67.00 \cite{he2024strong} \\
TaS$_2$-Sc$_2$NCl$_2$ & 23.37     & WSSiN$_2$ & 65.04 \cite{ma2025coexistence} \\
NbSe$_2$-ZrBr  & 22.30            & Na$_2$Te/Cs$_2$S & 59.68 \cite{arora2023negative} \\
TaSe$_2$-ZrBr  & 4.18             & CaFI & 50.68 \cite{ma2025origin} \\
\cline{1-2}
ZrBr-ZrS$_2$  & 60.17             & Y$_2$COSe, Y$_2$COS & 42.64, 49.86 \cite{yang2025large} \\
ZrBr-HfS$_2$  & 58.91             & LiInTe$_2$, LiAlS$_2$ & 12, 44 \cite{liu2021coexistence} \\
ZrBr-ZrNCl  & 49.56               & TiS$_2$Se & 43.58 \cite{ahammed2020ultrahigh} \\
NbS$_2$-FeCl$_2$  & 31.25         & TiSe$_2$S & 43.58 \cite{ahammed2020ultrahigh} \\
TaS$_2$-FeCl$_2$  & 30.27         & Ga$_2$SeTe & 29.0 \cite{guo2017enhanced} \\
TaSe$_2$-ZrCl$_2$  & 21.09        & $\gamma$-GeSnSeO & 25.94 \cite{zhang2024intrinsic} \\
NbS$_2$-ZrCl$_2$  & 17.45         & InSb, InP & 2.3, 25.1 \cite{blonsky2015ab} \\
ZrBr-SnS$_2$  & 16.86             & $\gamma$-GeSnSO & 20.01 \cite{zhang2024intrinsic} \\
\cline{1-2}
Cd(OH)$_2$-ZrS$_2$  & 83.35 (76.86)       & Sc$_2$CSO & 14.00 \cite{zhang2022large} \\ 
Sc$_2$CCl$_2$-Cd(OH)$_2$  & 70.75 (65.15) & In$_2$NO$_2$ & 13.90 \cite{ai2022ferroelectricity} \\ 
Tl$_2$O-SnS$_2$  & 69.53 (67.01)          & TiP & 13.00 \cite{li2025enhanced} \\
Cd(OH)$_2$-HfS$_2$  & 68.86 (58.86)       & GaAs, GaSb & 8.2, 0.8 \cite{blonsky2015ab} \\
ZrCl$_2$-Sc$_2$CCl$_2$ & 26.89 (24.76)    & Ga$_2$SSe & 8.0 \cite{guo2017enhanced} \\
Tl$_2$O-ZrS$_2$ & 18.47 (18.12)           & SiH-CdCl$_2$ & 7.75 \cite{priydarshi2023versatility} \\ 
Cd(OH)$_2$-SnS$_2$  &  9.27 (8.92)       & $\gamma$-LiGaS$_2$ & 3.22 \cite{lv2023large} \\
PtS$_2$-Cd(OH)$_2$  & 6.36 (8.35)      & $\beta$-SbAs, AsN, AsP & 1.4,15,1.0 \cite{yin2017giant} \\
\hline
\end{tabular}
}
\label{tab1}
\end{table}

We also obtained the ``effective biaxial electromechanical strain coefficient'' $d_{3,biax}$ by dividing the $e_{3,biax}$ values by ``2D biaxial elastic stiffness'' $C$. The 2D biaxial elastic stiffness values are calculated by fitting the energy-biaxial strain curve. Figure~\ref{fig5}(b) demonstrates a strong positive linear correlation between $e_{3,biax}$ and $d_{3,biax}$. Larger $e_{3,biax}$ reflects a stronger polarization response induced by mechanical deformation, which consequently enhances the strain generated under an electric field. Therefore, materials possessing larger piezoelectric stress coefficients generally exhibit higher piezoelectric strain coefficients also. The positive correlation reflects the dominant contribution of \(e_{3,\mathrm{biax}}\), while deviations from the trend arise from differences in the 2D biaxial elastic stiffness. Specifically, a material with lower 2D biaxial elastic stiffness requires less mechanical energy to deform and therefore exhibits larger $d_{3,biax}$. Thus, electromechanical performance requires a balance between high strain sensitivity of polarization and moderate elastic stiffness.

The relationship between the work function difference ($\Delta\Phi$) of the constituent monolayers and the polarization of the corresponding van der Waals HSs is illustrated in Figure~\ref{fig5}(c). When two monolayers possessing different work functions are brought into contact, electrons spontaneously transfer from the material with the lower work function to the one with the higher work function until their Fermi levels become aligned. Thus, the work function difference serves as the thermodynamic driving force for interfacial charge transfer during HS formation. This redistribution of charge generates an interfacial dipole and forms a built-in electric field across the heterointerface. As the work function difference increases, the amount of charge transferred often increases, resulting in a strong built-in electric field and improved segregation of positive and negative charges at the interface. Consequently, the dipole moment increases, leading to a larger spontaneous polarization in the HS. The overall positive correlation observed in Figure~\ref{fig5}(c) thus suggests that the work-function difference can serve as a useful electronic descriptor for estimating the magnitude of interfacial polarization in van der Waals HSs.

\section{Conclusion}
This work presents a high-throughput \textit{ab initio} study of the out-of-plane electromechanical response of 2D heterostructures by screening 461 HSs constructed from 57 monolayers. A central outcome of this work is the distinction between strain-tunable interfacial dipoles in metallic HSs and conventional piezoelectric responses in semiconducting HSs. In M--M and M--S systems, interface-induced charge redistribution produces finite out-of-plane dipoles that evolve systematically with strain, and we characterize these structures as polar metallic interfaces with tunable electromechanical dipoles. 

In S--S heterostructures, the same interfacial charge-transfer mechanism produces a well-defined macroscopic polarization, thereby enabling quantification of the conventional piezoelectric response using both CDD and Berry-phase methods. These results establish interfacial charge engineering as a common microscopic mechanism connecting polar metallic electromechanical interfaces and piezoelectric semiconducting heterostructures. 

Among the 24 selected candidates, the absolute out-of-plane biaxial electromechanical coefficient reaches a maximum of $e_{3,biax}=83.35$ pCm$^{-1}$, demonstrating that substantial electromechanical responses can arise from interface engineering. The identified heterostructures provide computationally predicted candidates for further investigation in nanoscale electromechanical and piezoelectric applications. More broadly, our results provide a design framework for identifying promising 2D piezoelectric materials through controlled symmetry breaking and interfacial charge engineering.

\section{Supporting Information}
Electronic band structures of the constituent monolayers, layer-resolved orbital-projected band structures of heterostructures, optimized crystal geometry of the heterostructures, Phonon dispersion, planar-average electrostatic potential, Bader charge analysis, and charge-transfer analysis of the heterostructures. 

\section{Acknowledgements}
We acknowledge the National Super Computing Mission (NSM) for providing computing resources of ``PARAM Sanganak'' at IIT Kanpur, which is implemented by CDAC and supported by the Ministry of Electronics and Information Technology (MeitY) and the Department of Science and Technology (DST), Government of India. The authors also acknowledge the High Performance Computing (HPC) facility PARAM Rudra at IUAC, New Delhi, for providing the computational resources used in this work.

\bibliography{ref}

@article{Grimme2010,
    author = {Grimme, Stefan and Antony, Jens and Ehrlich, Stephan and Krieg, Helge},
    title = {A consistent and accurate ab initio parametrization of density functional dispersion correction (DFT-D) for the 94 elements H-Pu},
    journal = {The Journal of Chemical Physics},
    volume = {132},
    number = {15},
    pages = {154104},
    year = {2010},
    month = {04},
    issn = {0021-9606},
    doi = {10.1063/1.3382344},
    url = {https://doi.org/10.1063/1.3382344},
    eprint = {https://pubs.aip.org/aip/jcp/article-pdf/doi/10.1063/1.3382344/15684000/154104_1_online.pdf},
}

@article{king1993,
  title = {Theory of polarization of crystalline solids},
  author = {King-Smith, R. D. and Vanderbilt, David},
  journal = {Phys. Rev. B},
  volume = {47},
  issue = {3},
  pages = {1651(R)--1654(R)},
  numpages = {0},
  year = {1993},
  month = {Jan},
  publisher = {American Physical Society},
  doi = {10.1103/PhysRevB.47.1651},
  url = {https://link.aps.org/doi/10.1103/PhysRevB.47.1651}
}

@article{10.1021/nn1029789,
    author = {Dresselhaus, Mildred S. and Araujo, Paulo T.},
    title = {Perspectives on the 2010 Nobel Prize in Physics for Graphene},
    journal = {ACS Nano},
    volume = {4},
    number = {11},
    pages = {6297-6302},
    year = {2010},
    month = {11},
    issn = {1936-0851},
    doi = {10.1021/nn1029789},
    url = {https://doi.org/10.1021/nn1029789},
    eprint = {https://pubs.acs.org/ancac3/article-pdf/4/11/6297/16052123/nn1029789.pdf},
}

@article{PhysRevB.102.035420,
  title = {${\mathrm{K}}_{2}{\mathrm{CoS}}_{2}$: A two-dimensional in-plane antiferromagnetic insulator},
  author = {Sarkar, Anan Bari and Ghosh, Barun and Singh, Bahadur and Bhowmick, Somnath and Lin, Hsin and Bansil, Arun and Agarwal, Amit},
  journal = {Phys. Rev. B},
  volume = {102},
  issue = {3},
  pages = {035420},
  numpages = {8},
  year = {2020},
  month = {Jul},
  publisher = {American Physical Society},
  doi = {10.1103/PhysRevB.102.035420},
  url = {https://link.aps.org/doi/10.1103/PhysRevB.102.035420}
}

@article{puri2018,
  title = {External-strain-induced semimetallic and metallic phase of chlorographene},
  author = {Puri, Shivam and Bhowmick, Somnath},
  journal = {Phys. Rev. Mater.},
  volume = {2},
  issue = {4},
  pages = {044001},
  numpages = {6},
  year = {2018},
  month = {Apr},
  publisher = {American Physical Society},
  doi = {10.1103/PhysRevMaterials.2.044001},
  url = {https://link.aps.org/doi/10.1103/PhysRevMaterials.2.044001}
}

@article{mounet2018two,
  title={Two-dimensional materials from high-throughput computational exfoliation of experimentally known compounds},
  author={Mounet, Nicolas and Gibertini, Marco and Schwaller, Philippe and Campi, Davide and Merkys, Andrius and Marrazzo, Antimo and Sohier, Thibault and Castelli, Ivano Eligio and Cepellotti, Andrea and Pizzi, Giovanni and others},
  journal={Nature nanotechnology},
  volume={13},
  number={3},
  pages={246--252},
  year={2018},
  publisher={Nature Publishing Group UK London},
doi={10.1038/s41565-017-0035-5},
URL={https://doi.org/10.1038/s41565-017-0035-5}
}

@article{blonsky2015ab,
  title={Ab initio prediction of piezoelectricity in two-dimensional materials},
  author={Blonsky, Michael N and Zhuang, Houlong L and Singh, Arunima K and Hennig, Richard G},
  journal={ACS nano},
  volume={9},
  number={10},
  pages={9885--9891},
  year={2015},
  publisher={ACS Publications},
  doi = {10.1021/acsnano.5b03394},
  URL = {https://doi.org/10.1021/acsnano.5b03394}
}

@article{zelisko2014anomalous,
  title={Anomalous piezoelectricity in two-dimensional graphene nitride nanosheets},
  author={Zelisko, Matthew and Hanlumyuang, Yuranan and Yang, Shubin and Liu, Yuanming and Lei, Chihou and Li, Jiangyu and Ajayan, Pulickel M and Sharma, Pradeep},
  journal={Nature communications},
  volume={5},
  number={1},
  pages={4284},
  year={2014},
  publisher={Nature Publishing Group UK London},
  DOI  = {10.1038/ncomms5284},
  URL  = {https://doi.org/10.1038/ncomms5284}
}

@article{duerloo2012intrinsic,
  title={Intrinsic piezoelectricity in two-dimensional materials},
  author={Duerloo, Karel-Alexander N and Ong, Mitchell T and Reed, Evan J},
  journal={The Journal of Physical Chemistry Letters},
  volume={3},
  number={19},
  pages={2871--2876},
  year={2012},
  publisher={ACS Publications},
  doi = {10.1021/jz3012436},
  URL = { https://doi.org/10.1021/jz3012436}
}

@article{giannozzi2009quantum,
  title={QUANTUM ESPRESSO: a modular and open-source software project for quantum simulations of materials},
  author={Giannozzi, Paolo and Baroni, Stefano and Bonini, Nicola and Calandra, Matteo and Car, Roberto and Cavazzoni, Carlo and Ceresoli, Davide and Chiarotti, Guido L and Cococcioni, Matteo and Dabo, Ismaila and others},
  journal={Journal of physics: Condensed matter},
  volume={21},
  number={39},
  pages={395502},
  year={2009},
  publisher={IOP Publishing},
doi  ="10.1088/0953-8984/21/39/395502",
url  ="https://dx.doi.org/10.1088/0953-8984/21/39/395502",
}

@article{PhysRevLett.77.3865,
  title = {Generalized Gradient Approximation Made Simple},
  author = {Perdew, John P. and Burke, Kieron and Ernzerhof, Matthias},
  journal = {Phys. Rev. Lett.},
  volume = {77},
  issue = {18},
  pages = {3865--3868},
  numpages = {0},
  year = {1996},
  month = {Oct},
  publisher = {American Physical Society},
  doi = {10.1103/PhysRevLett.77.3865},
  url = {https://link.aps.org/doi/10.1103/PhysRevLett.77.3865}
}

@article{PhysRevB.50.17953,
  title = {Projector augmented-wave method},
  author = {Bl\"ochl, P. E.},
  journal = {Phys. Rev. B},
  volume = {50},
  issue = {24},
  pages = {17953--17979},
  numpages = {0},
  year = {1994},
  month = {Dec},
  publisher = {American Physical Society},
  doi = {10.1103/PhysRevB.50.17953},
  url = {https://link.aps.org/doi/10.1103/PhysRevB.50.17953}
}

@article{momma2008vesta,
  title={VESTA: a three-dimensional visualization system for electronic and structural analysis},
  author={Momma, Koichi and Izumi, Fujio},
  journal={Journal of Applied crystallography},
  volume={41},
  number={3},
  pages={653--658},
  year={2008},
  publisher={International Union of Crystallography},
doi = {10.1107/S0021889808012016},
  url = {https://doi.org/10.1107/S0021889808012016}
}

@article{pham2021enhanced,
  title={Enhanced out-of-plane electromechanical response of Janus {ZrSeO}},
  author={Pham, Thi Hue and Ullah, Hamid and Shafique, Aamir and Kim, Hye Jung and Shin, Young-Han},
  journal={Physical Chemistry Chemical Physics},
  volume={23},
  number={30},
  pages={16289--16295},
  year={2021},
  publisher={Royal Society of Chemistry},
  doi  ={10.1039/D1CP00119A},
url  ={http://dx.doi.org/10.1039/D1CP00119A}
}

@article{mahajan2023magnetoelectric,
  title={Magnetoelectric multiferroic Janus monolayers {VOXY (X/Y= F, Cl, Br, or I, and X$\ne$ Y)} with in-plane ferroelectricity and out-of-plane piezoelectricity},
  author={Mahajan, Akshay and Bhowmick, Somnath},
  journal={The Journal of Physical Chemistry C},
  volume={127},
  number={23},
  pages={11407--11418},
  year={2023},
  publisher={ACS Publications},
  doi = {10.1021/acs.jpcc.3c03100},
URL = {https://doi.org/10.1021/acs.jpcc.3c03100}
}

@article{zhang2025promising,
  title={A promising direct Z-scheme {GaAs/HfS$_2$} van der Waals heterostructure photocatalyst for overall water-splitting using infrared, visible and ultraviolet solar spectrum},
  author={Zhang, Yan and Li, Jia-Hui and Duan, Li},
  journal={Chemical Physics},
  pages={112965},
  year={2025},
  publisher={Elsevier},
  doi = {https://doi.org/10.1016/j.chemphys.2025.112965},
url = {https://www.sciencedirect.com/science/article/pii/S0301010425003660}
}

@article{xie2026modulating,
  title={Modulating the contact properties of {XY$_2$/Sc$_2$ CCl$_2$ (X= Nb, Ni, Ti, V, Mn, Ta; Y= S, Se)} heterostructures via layer number, electric field, and vertical strain},
  author={Xie, You and Zhang, Miao and Chen, Zheng-Yong and Du, Chen and Li, Jia-Qi and Chen, Yan and Hao, Li-Mei and Zhang, Tao},
  journal={Physical Chemistry Chemical Physics},
  year={2026},
  publisher={Royal Society of Chemistry},
  doi  ="10.1039/D5CP04345J",
url  ="http://dx.doi.org/10.1039/D5CP04345J"
}

@article{wang2019modulating,
  title={Modulating charge density wave order in a {1T-TaS$_2$}/black phosphorus heterostructure},
  author={Wang, Ziying and Chu, Leiqiang and Li, Linjun and Yang, Ming and Wang, Junyong and Eda, Goki and Loh, Kian Ping},
  journal={Nano Letters},
  volume={19},
  number={5},
  pages={2840--2849},
  year={2019},
  publisher={ACS Publications},
  doi = {10.1021/acs.nanolett.8b04805},
  URL = {https://doi.org/10.1021/acs.nanolett.8b04805}
}

@article{khang2024achieving,
  title={Achieving ohmic contacts in {NbS$_2$/MoSe$_2$ van der Waals} heterostructure: a first-principles study},
  author={Khang, Nguyen D and Nguyen, Cuong Q and Nguyen, Chuong V},
  journal={Advanced Theory and Simulations},
  volume={7},
  number={3},
  pages={2300757},
  year={2024},
  publisher={Wiley Online Library},
  doi = {https://doi.org/10.1002/adts.202300757},
url = {https://advanced.onlinelibrary.wiley.com/doi/abs/10.1002/adts.202300757}
}

@article{yang2019coexistence,
  title={Coexistence of piezoelectricity and magnetism in two-dimensional vanadium dichalcogenides},
  author={Yang, Jianhui and Wang, Anping and Zhang, Shaozheng and Liu, Jia and Zhong, Zhicheng and Chen, Liang},
  journal={Physical Chemistry Chemical Physics},
  volume={21},
  number={1},
  pages={132--136},
  year={2019},
  publisher={Royal Society of Chemistry},
  doi  ={"10.1039/C8CP06535G"},
url  ={"http://dx.doi.org/10.1039/C8CP06535G"}
}

@article{ahammed2020ultrahigh,
  title={Ultrahigh out-of-plane piezoelectricity meets giant Rashba effect in {2D} Janus monolayers and bilayers of group {IV} transition-metal trichalcogenides},
  author={Ahammed, Raihan and Jena, Nityasagar and Rawat, Ashima and Mohanta, Manish K and Dimple and De Sarkar, Abir},
  journal={The Journal of Physical Chemistry C},
  volume={124},
  number={39},
  pages={21250--21260},
  year={2020},
  publisher={ACS Publications},
  URL = {https://doi.org/10.1021/acs.jpcc.0c05134}
}

@article{ai2022ferroelectricity,
  title={Ferroelectricity coexisted with p-orbital ferromagnetism and metallicity in two-dimensional metal oxynitrides},
  author={Ai, Haoqiang and Li, Feifei and Bai, Haoyun and Liu, Dong and Lo, Kin Ho and Yang, Shengyuan A and Kawazoe, Yoshiyuki and Pan, Hui},
  journal={npj Computational Materials},
  volume={8},
  number={1},
  pages={60},
  year={2022},
  publisher={Nature Publishing Group UK London},
  DOI  = {10.1038/s41524-022-00737-3},
URL  ={ https://doi.org/10.1038/s41524-022-00737-3}

}

@article{yang2025large,
  title={Large out-of-plane piezoelectricity in {MXene Y$_2$COX (X= O, S, Se)} monolayers},
  author={Yang, Hao-Hao and Zhao, Jiao and Cao, Yi-Ming and Kang, Yan-Ru and Zheng, Shi-Yun and Fan, Xin and Li, Han-Wei and Liu, Hong-Wei and Wei, Sheng-Xian and Chen, Qi and others},
  journal={Surfaces and Interfaces},
  volume={58},
  pages={105834},
  year={2025},
  publisher={Elsevier},
  doi = {https://doi.org/10.1016/j.surfin.2025.105834},
url = {https://www.sciencedirect.com/science/article/pii/S2468023025000975}
}

@article{yin2023monolayer,
  title={Monolayer polar metals with large piezoelectricity derived from {MoSi$_2$ N$_4$}},
  author={Yin, Yan and Gong, Qihua and Yi, Min and Guo, Wanlin},
  journal={Materials Horizons},
  volume={10},
  number={11},
  pages={5177--5184},
  year={2023},
  publisher={Royal Society of Chemistry},
  doi  ={"10.1039/D3MH00743J"},
url  ={"http://dx.doi.org/10.1039/D3MH00743J"}
}

@article{arora2023negative,
  title={Negative piezoelectricity and enhanced electrical conductivity at the interfaces of two-dimensional dialkali oxide and chalcogenide monolayers},
  author={Arora, Anu and Rawat, Ashima and De Sarkar, Abir},
  journal={Physical Review B},
  volume={107},
  number={8},
  pages={085402},
  year={2023},
  publisher={APS},
    doi = {10.1103/PhysRevB.107.085402},
  url = {https://link.aps.org/doi/10.1103/PhysRevB.107.085402}
}

@article{jia2018piezoelectric,
  title={Piezoelectric and polarized enhancement by hydrofluorination of penta-graphene},
  author={Jia, Hao-Jun and Mu, Hui-Min and Li, Jin-Peng and Zhao, Yan-Zhen and Wu, Yu-Xuan and Wang, Xiao-Chun},
  journal={Physical Chemistry Chemical Physics},
  volume={20},
  number={41},
  pages={26288--26296},
  year={2018},
  publisher={Royal Society of Chemistry},
  doi  ={"10.1039/C8CP04010A"},
url  ={"http://dx.doi.org/10.1039/C8CP04010A"}
}

@article{li2025enhanced,
  title={Enhanced out-of-plane piezoelectricity in two-dimensional {TlP} by surface hydrogenation},
  author={Li, Jia-Bin and Tang, Yonghua},
  journal={AIP Advances},
  volume={15},
  number={3},
  year={2025},
  publisher={AIP Publishing},
  doi = {10.1063/5.0240846},
    url = {https://doi.org/10.1063/5.0240846}
}

@article{liu2021coexistence,
  title={Coexistence of large out-of-plane and in-plane piezoelectricity in {2D monolayer Li-based ternary chalcogenides LiMX$_2$}},
  author={Liu, Siyuan and Chen, Weizhen and Liu, Chang and Wang, Bing and Yin, Huabing},
  journal={Results in Physics},
  volume={26},
  pages={104398},
  year={2021},
  publisher={Elsevier},
  doi = {https://doi.org/10.1016/j.rinp.2021.104398},
url = {https://www.sciencedirect.com/science/article/pii/S2211379721005209}
}

@article{ma2025origin,
  title={Origin of Negative Longitudinal Piezoelectricity Based on {2D bipolar OOP} Piezoelectric Metal Halides},
  author={Ma, Xikui and Guo, Yiyi and Fan, Yingcai and Bu, Hongxia and Zhao, Mingwen},
  journal={ACS omega},
  volume={10},
  number={39},
  pages={45410--45416},
  year={2025},
  publisher={ACS Publications},
  doi = {10.1021/acsomega.5c05327},
URL = {https://doi.org/10.1021/acsomega.5c05327}
}

@article{he2024strong,
  title={Strong vertical piezoelectricity and broad-pH-value photocatalyst in ferroelastic {Y$_2$Se$_2$BrF} monolayer},
  author={He, Qi-Wen and Wang, Jun-Hui and Zhu, Dan-Yang and Tang, Dai-Song and Lv, Zengtao and Guo, Feng and Wang, Xiao-Chun},
  journal={Nano Letters},
  volume={24},
  number={29},
  pages={8979--8987},
  year={2024},
  publisher={ACS Publications},
  doi = {10.1021/acs.nanolett.4c01943},
  URL = {https://doi.org/10.1021/acs.nanolett.4c01943}
}

@article{priydarshi2023versatility,
  title={Versatility of the {SiH--CdCl$_2$} heterostructure: piezoelectricity, photocatalysis, and transistor applications},
  author={Priydarshi, Achintya and Arora, Abhinav and Chauhan, Yogesh Singh and Agarwal, Amit and Bhowmick, Somnath},
  journal={The Journal of Physical Chemistry C},
  volume={127},
  number={43},
  pages={21279--21288},
  year={2023},
  publisher={ACS Publications},
  doi = {10.1021/acs.jpcc.3c04021},
URL = {https://doi.org/10.1021/acs.jpcc.3c04021}
}

@article{ma2021large,
  title={Large family of two-dimensional ferroelectric metals discovered via machine learning},
  author={Ma, Xing-Yu and Lyu, Hou-Yi and Hao, Kuan-Rong and Zhao, Yi-Ming and Qian, Xiaofeng and Yan, Qing-Bo and Su, Gang},
  journal={Science Bulletin},
  volume={66},
  number={3},
  pages={233--242},
  year={2021},
  publisher={Elsevier},
  doi = {https://doi.org/10.1016/j.scib.2020.09.010},
url = {https://www.sciencedirect.com/science/article/pii/S2095927320305892}
}

@article{tang2021first,
  title={First-principles prediction of polar half-metallicity and out-of-plane piezoelectricity in two-dimensional quintuple layered cobalt selenide},
  author={Tang, Cheng and Zhang, Lei and Jiao, Yalong and Zhang, Chunmei and Sanvito, Stefano and Du, Aijun},
  journal={Journal of Materials Chemistry C},
  volume={9},
  number={36},
  pages={12046--12050},
  year={2021},
  publisher={Royal Society of Chemistry},
  doi  ={"10.1039/D1TC02727A"},
url  ={"http://dx.doi.org/10.1039/D1TC02727A"}
}

@article{zhang2020first,
  title={First-principles screening of novel ferroelectric MXene phases with a large piezoelectric response and unusual auxeticity},
  author={Zhang, Lei and Tang, Cheng and Zhang, Chunmei and Du, Aijun},
  journal={Nanoscale},
  volume={12},
  number={41},
  pages={21291--21298},
  year={2020},
  publisher={Royal Society of Chemistry},
  doi  ={"10.1039/D0NR06609E"},
url  ={"http://dx.doi.org/10.1039/D0NR06609E"}
}

@article{naseer2024room,
  title={Room temperature ferroelectricity and an electrically tunable Berry curvature dipole in {III--V} monolayers},
  author={Naseer, Ateeb and Priydarshi, Achintya and Ghosh, Pritam and Ahammed, Raihan and Chauhan, Yogesh Singh and Bhowmick, Somnath and Agarwal, Amit},
  journal={Nanoscale},
  volume={16},
  number={25},
  pages={12107--12117},
  year={2024},
  publisher={Royal Society of Chemistry},
  doi  ={"10.1039/D4NR00336E"},
url  ={"http://dx.doi.org/10.1039/D4NR00336E"}
}

@article{zhang2019first,
  title={First-principles prediction of a room-temperature ferromagnetic {Janus VSSe} monolayer with piezoelectricity, ferroelasticity, and large valley polarization},
  author={Zhang, Chunmei and Nie, Yihan and Sanvito, Stefano and Du, Aijun},
  journal={Nano letters},
  volume={19},
  number={2},
  pages={1366--1370},
  year={2019},
  publisher={ACS Publications},
  doi = {10.1021/acs.nanolett.8b05050},
  URL = {https://doi.org/10.1021/acs.nanolett.8b05050}
}

@article{gan2024emergence,
  title={Emergence of In-Plane and Out-of-Plane Piezoelectricity in Flexible {Janus MoXSiN$_2$ (X = S, Se, and Te)} Monolayers},
  author={Gan, Wantao and Ma, Xinguo and Liao, Jiajun and Xie, Tian and Ma, Nan and Lu, Jingjing and Hong, Jeongmin},
  journal={ACS Applied Electronic Materials},
  volume={6},
  number={8},
  pages={5951--5960},
  year={2024},
  publisher={ACS Publications},
  doi = {10.1021/acsaelm.4c00894},
  url = {https://doi.org/10.1021/acsaelm.4c00894}
}

@article{bezzerga2023ferro,
  title={Ferro-piezoelectricity in emerging {Janus monolayer BMX$_2$ (M= Ga, In and X= S, Se)}: ab initio investigations},
  author={Bezzerga, Djamel and Haidar, El-Abed and Stampfl, Catherine and Mir, Ali and Sahnoun, Mohammed},
  journal={Nanoscale Advances},
  volume={5},
  number={5},
  pages={1425--1432},
  year={2023},
  publisher={Royal Society of Chemistry},
  doi  ={"10.1039/D2NA00597B"},
url  ={"http://dx.doi.org/10.1039/D2NA00597B"}
}

@article{dong2017large,
  title={Large in-plane and vertical piezoelectricity in {Janus} transition metal dichalchogenides},
  author={Dong, Liang and Lou, Jun and Shenoy, Vivek B},
  journal={ACS nano},
  volume={11},
  number={8},
  pages={8242--8248},
  year={2017},
  publisher={ACS Publications},
  doi = {10.1021/acsnano.7b03313},
  URL = { https://doi.org/10.1021/acsnano.7b03313}
}

@article{zhu2015observation,
  title={Observation of piezoelectricity in free-standing monolayer {MoS$_2$}},
  author={Zhu, Hanyu and Wang, Yuan and Xiao, Jun and Liu, Ming and Xiong, Shaomin and Wong, Zi Jing and Ye, Ziliang and Ye, Yu and Yin, Xiaobo and Zhang, Xiang},
  journal={Nature nanotechnology},
  volume={10},
  number={2},
  pages={151--155},
  year={2015},
  publisher={Nature Publishing Group UK London},
  DOI  ={ 10.1038/nnano.2014.309},
  URL  = {https://doi.org/10.1038/nnano.2014.309}
}

@article{novoselov20162d,
  title={{2D materials and van der Waals} heterostructures},
  author={Novoselov, K Sꎬ and Mishchenko, Artem and Carvalho, Alexandra and Castro Neto, AH},
  journal={Science},
  volume={353},
  number={6298},
  pages={aac9439},
  year={2016},
  publisher={American Association for the Advancement of Science},
  doi = {10.1126/science.aac9439},
URL = {https://www.science.org/doi/abs/10.1126/science.aac9439}
}

@article{wang2006piezoelectric,
  title={Piezoelectric field effect transistor and nanoforce sensor based on a single {ZnO} nanowire},
  author={Wang, Xudong and Zhou, Jun and Song, Jinhui and Liu, Jin and Xu, Ningsheng and Wang, Zhong L},
  journal={Nano letters},
  volume={6},
  number={12},
  pages={2768--2772},
  year={2006},
  publisher={ACS Publications},
  doi = {10.1021/nl061802g},
  URL = {https://doi.org/10.1021/nl061802g}
}

@article{ong2012engineered,
  title={Engineered piezoelectricity in graphene},
  author={Ong, Mitchell T and Reed, Evan J},
  journal={ACS nano},
  volume={6},
  number={2},
  pages={1387--1394},
  year={2012},
  publisher={ACS Publications},
  doi = {10.1021/nn204198g},
  URL = {https://doi.org/10.1021/nn204198g}
}

@article{yadav2025high,
  title={High-Throughput Screening of {2D} Photocatalyst Heterostructures with Suppressed Electron--Hole Recombination for Solar Water Splitting},
  author={Yadav, Shivanand and Modi, Jainandan Kumar and Ahammed, Raihan and Bhadoria, BS and Chauhan, Yogesh S and Agarwal, Amit and Bhowmick, Somnath},
  journal={ACS Applied Energy Materials},
  volume={8},
  number={19},
  pages={14601--14611},
  year={2025},
  publisher={ACS Publications},
  doi = {10.1021/acsaem.5c02326},
URL = { https://doi.org/10.1021/acsaem.5c02326}
}

@article{li2023van,
  title={A van der Waals heterostructure with an electronically textured moir{\'e} pattern: {PtSe$_2$/PtTe$_2$}},
  author={Li, Jingfeng and Ghorbani-Asl, Mahdi and Lasek, Kinga and Pathirage, Vimukthi and Krasheninnikov, Arkady V and Batzill, Matthias},
  journal={ACS nano},
  volume={17},
  number={6},
  pages={5913--5920},
  year={2023},
  doi = {10.1021/acsnano.2c12879},
  URL = {https://doi.org/10.1021/acsnano.2c12879},
  publisher={ACS Publications}
}

@article{rizu2025uv,
  title={{UV}-enhanced exfoliated {MoS$_2$/PtSe$_2$} heterostructure for ultra-sensitive {NO$_2$} detection at room temperature},
  author={Rizu, Mubdiul Islam and Fadil, Dalal and Llobet, Eduard},
  journal={npj 2D Materials and Applications},
  volume={9},
  number={1},
  pages={28},
  year={2025},
  DO  = {10.1038/s41699-025-00548-2},
  URL  = {https://doi.org/10.1038/s41699-025-00548-2},
  publisher={Nature Publishing Group UK London}
}

@article{ma2025coexistence,
  title={Coexistence of in-plane and out-of-plane piezoelectricity in {Janus WXSiN$_2$ (X= S, Se, Te)} monolayers: potential implications for nanoscale piezoelectric devices},
  author={Ma, Biao and Cui, Shou-Xin and Zhao, Bao and Zhang, Guiqing and Guo, Feng and Li, Jun and Wang, Xiao-Chun},
  journal={ACS Applied Nano Materials},
  volume={8},
  number={9},
  pages={4876--4884},
  year={2025},
  doi = {10.1021/acsanm.5c00386},
 URL = {https://doi.org/10.1021/acsanm.5c00386},
  publisher={ACS Publications}
}

@article{zhang2024intrinsic,
  title={Intrinsic coupling between piezoelectric and electronic transport properties in {Janus $\gamma$-GeSnXO (X= S, Se)} monolayers with vertical piezoelectricity},
  author={Zhang, He-Na and Shang, Xiao and He, Qi-Wen and Tang, Dai-Song and Wang, Jun-Hui and Zhu, Dan-Yang and Wang, Xiao-Chun},
  journal={ACS Applied Nano Materials},
  volume={7},
  number={8},
  pages={8969--8977},
  year={2024},
  doi = {10.1021/acsanm.4c00339},
  URL = {https://doi.org/10.1021/acsanm.4c00339},
  publisher={ACS Publications}
}

@article{li2024intrinsically,
  title={Intrinsically asymmetric atomic character regulates piezoelectricity in two-dimensional materials},
  author={Li, Yun-Qin and He, Qi-Wen and Tang, Dai-Song and Shang, Xiao and Wang, Xiao-Chun},
  journal={Frontiers of Physics},
  volume={19},
  number={3},
  pages={33201},
  year={2024},
  URL = {https://doi.org/10.1007/s11467-023-1348-5},
  DOI  = {10.1007/s11467-023-1348-5},
  publisher={Springer}
}

@article{zhang2022large,
  title={Large out-of-plane piezoelectricity of {VIA group functionalized MXenes thin films for MEMS}},
  author={Zhang, He-Na and Yang, Chunhua and Li, Yun-Qin and Zhu, Shi-Yu and Wang, Xin-Yu and He, Qi-Wen and Tang, Dai-Song and Wang, Xiao-Chun},
  journal={Applied Physics Letters},
  volume={121},
  number={14},
  year={2022},
  doi = {10.1063/5.0106898},
 url = {https://doi.org/10.1063/5.0106898},
  publisher={AIP Publishing}
}

@article{yin2017giant,
  title={Giant piezoelectric effects in monolayer group-{V} binary compounds with honeycomb phases: a first-principles prediction},
  author={Yin, Huabing and Gao, Jingwei and Zheng, Guang-Ping and Wang, Yuanxu and Ma, Yuchen},
  journal={The Journal of Physical Chemistry C},
  volume={121},
  number={45},
  pages={25576--25584},
  year={2017},
  doi = {10.1021/acs.jpcc.7b08822},
 URL = {https://doi.org/10.1021/acs.jpcc.7b08822},
  publisher={ACS Publications}
}

@article{guo2017enhanced,
  title={Enhanced piezoelectric effect in {Janus group-III} chalcogenide monolayers},
  author={Guo, Yu and Zhou, Si and Bai, Yizhen and Zhao, Jijun},
  journal={Applied Physics Letters},
  volume={110},
  number={16},
  year={2017},
  doi = {10.1063/1.4981877},
    url = {https://doi.org/10.1063/1.4981877},
  publisher={AIP Publishing}
}

@article{lv2023large,
  title={Large in-plane and out-of-plane piezoelectricity in {2D $\gamma$-LiMX$_2$ (M= Al, Ga and In; X= S, Se and Te)} monolayers},
  author={Lv, Qiaoya and Qiu, Jian and Wen, Quan and Li, Dongling and Zhou, Yuanyuan and Lu, Guofeng},
  journal={Materials Science in Semiconductor Processing},
  volume={154},
  pages={107222},
  year={2023},
  doi = {https://doi.org/10.1016/j.mssp.2022.107222},
url = {https://www.sciencedirect.com/science/article/pii/S136980012200748X},
  publisher={Elsevier}
}

@article{yu2018piezoelectricity,
  title={Piezoelectricity in {WSe$_2$/MoS$_2$} heterostructure atomic layers},
  author={Yu, Sheng and Rice, Quinton and Tabibi, Bagher and Li, Qiliang and Seo, Felix Jaetae},
  journal={Nanoscale},
  volume={10},
  number={26},
  pages={12472--12479},
  year={2018},
  doi = {10.1039/c8nr04394a},
  url = {https://doi.org/10.1039/c8nr04394a},
  publisher={The Royal Society of Chemistry}
}

@article{mohanta2019superhigh,
  title={Superhigh out-of-plane piezoelectricity, low thermal conductivity and photocatalytic abilities in ultrathin 2D van der Waals heterostructures of boron monophosphide and gallium nitride},
  author={Mohanta, Manish Kumar and Rawat, Ashima and Dimple and Jena, Nityasagar and Ahammed, Raihan and De Sarkar, Abir},
  journal={Nanoscale},
  volume={11},
  number={45},
  pages={21880--21890},
  year={2019},
  publisher={The Royal Society of Chemistry}
}

@article{yuan2020enhanced,
  title={Enhanced piezoelectric response of layered {In$_2$Se$_3$/MoS$_2$} nanosheet-based van der Waals heterostructures},
  author={Yuan, Shuoguo and Io, Weng Fu and Mao, Jianfeng and Chen, Yancong and Luo, Xin and Hao, Jianhua},
  journal={ACS Applied Nano Materials},
  volume={3},
  number={12},
  pages={11979--11986},
  year={2020},
  doi = {10.1021/acsanm.0c02513},
  url = {https://doi.org/10.1021/acsanm.0c02513},
  publisher={ACS Publications}
}

@article{chen2021multilayered,
  title={Multilayered vertical heterostructures comprised of {MoS$_2$} and {WS$_2$} nanosheets for optoelectronics},
  author={Chen, Fei and Jiang, Xia and Shao, Jiaqi and Zhao, Shichao and Su, Weitao and Li, Yajie},
  journal={ACS Applied Nano Materials},
  volume={4},
  number={9},
  pages={9293--9302},
  year={2021},
  doi = {10.1021/acsanm.1c01809},
  url = {https://doi.org/10.1021/acsanm.1c01809},
  publisher={ACS Publications}
}

@article{xu2020two,
  title={Two-dimensional polar metal of a PbTe monolayer by electrostatic doping},
  author={Xu, Tao and Zhang, Jingtong and Zhu, Yuquan and Wang, Jie and Shimada, Takahiro and Kitamura, Takayuki and Zhang, Tong-Yi},
  journal={Nanoscale Horizons},
  volume={5},
  number={10},
  pages={1400--1406},
  year={2020},
  doi = {10.1039/d0nh00188k},
    url = {https://doi.org/10.1039/d0nh00188k},
  publisher={The Royal Society of Chemistry}
}

@article{resta2002insulators,
  title={Why are insulators insulating and metals conducting?},
  author={Resta, Raffaele},
  journal={Journal of Physics: Condensed Matter},
  volume={14},
  number={20},
  pages={R625--R656},
  year={2002},
  doi = {10.1088/0953-8984/14/20/201},
url = {https://doi.org/10.1088/0953-8984/14/20/201}
}
\end{document}